\documentclass[fleqn,usenatbib]{mnras}
\usepackage{newtxtext,newtxmath}
\usepackage{enumitem} % FJ, for itemization without bullets and indents
\usepackage{soul} % FJ: for striking through text to be deleted 
\usepackage[T1]{fontenc}
\DeclareRobustCommand{\VAN}[3]{#2}
\let\VANthebibliography\thebibliography
\def\thebibliography{\DeclareRobustCommand{\VAN}[3]{##3}\VANthebibliography}
\usepackage{subfig}

\newcommand{\eq}[1]{eq.~(\ref{eq:#1})}

\newcommand{\se}[1]{Section \ref{sec:#1}}
\newcommand{\app}[1]{Appendix \ref{app:#1}}
\newcommand{\Fig}[1]{Fig.~\ref{fig:#1}}
\newcommand{\Figs}[1]{Figs.~\ref{fig:#1}}
\newcommand{\Tab}[1]{Table~\ref{tab:#1}}
\newcommand{\be}{\begin{equation}}
\newcommand{\ee}{\end{equation}}
\newcommand{\bad}{\begin{equation} \begin{aligned}}
\newcommand{\ead}{\end{aligned} \end{equation}}

\newcommand{\Msun}{M_\odot}

\newcommand{\Mv}{M_{\rm vir}}
\newcommand{\Ms}{M_{\star}}

\newcommand{\Mstar}{M_{\star}}

\newcommand{\Rv}{R_{\rm vir}}

\newcommand{\rhalf}{r_{\rm 1/2}}

\newcommand{\rmax}{r_{\rm max}}

\newcommand{\rc}{r_{\rm c}}

\newcommand{\Vc}{V_{\rm circ}}

\newcommand{\Vv}{V_{\rm vir}}

\newcommand{\Vmax}{V_{\rm max}}

\newcommand{\jc}{j_{\rm circ}}
\newcommand{\jz}{j_z}
\newcommand{\jp}{j_{\rm p}}
\newcommand{\jv}{j_{\rm vir}}

\newcommand{\lambdath}{\lambda_{\rm th}}

\newcommand{\dd}{\text{d}}
\newcommand{\etacut}{\eta_{\rm cut}}
\newcommand{\Phic}{\Phi_{\rm circ}}
\newcommand{\Ec}{E_{\rm circ}}
\newcommand{\eb}{e_{\rm b}}
\newcommand{\ecut}{e_{\rm cut}}
\newcommand{\fdisc}{f_{\rm disc}}
\newcommand{\fbulge}{f_{\rm bulge}}
\newcommand{\fhalo}{f_{\rm halo}}
\newcommand{\fthin}{f_{\rm thin}}
\newcommand{\fthick}{f_{\rm thick}}
\usepackage{float}
\usepackage{graphicx}	% Including figure files
\usepackage{amsmath}	% Advanced maths commands
\usepackage{bm}
\usepackage{array}
\newcolumntype{P}[1]{>{\arraybackslash}p{#1}}
\title[Galaxy-Halo Structural Connection I]
{Connection between galaxy morphology and dark-matter halo structure I: a running threshold for thin discs and size predictors from the dark sector }

\author[Jinning Liang et al.]{
Jinning Liang$^{1,2}$, 
Fangzhou Jiang$^{1}$\thanks{Corresponding author: fangzhou.jiang@pku.edu.cn}, 
Houjun Mo$^{3,4}$, 
Andrew Benson$^{5}$, 
Avishai Dekel$^{6}$, 
Noa Tavron$^{6}$, 
\newauthor
\,Philip F. Hopkins$^{7}$, 
Luis C. Ho$^{1,8}$
\\
$^{1}$Kavli Institute for Astronomy and Astrophysics, Peking University, Beijing 100871, China\\
$^{2}$Institute for Computational Cosmology, Department of Physics, Durham University, South Road, Durham DH1 3LE, UK\\
$^{3}$Department of Astronomy, University of Massachusetts, Amherst, MA01003, USA\\
$^{4}$Tsung-Dao Lee Institute, Shanghai Jiao Tong University, Shanghai 200240, China\\
$^{5}$ Carnegie Observatories, 813 Santa Barbara Street, Pasadena, CA 91101, USA\\
$^{6}$Center for Astrophysics and Planetary Science, Racah Institute of Physics, The Hebrew University, Jerusalem 91904, Israel\\
$^{7}$ TAPIR, California Institute of Technology, Pasadena, CA 91125, USA\\
$^{8}$ Department of Astronomy, School of Physics, Peking University, Beijing 100871, China
}

\date{Accepted XXX. Received YYY; in original form ZZZ}

\begin{document}
\label{firstpage}
\pagerange{\pageref{firstpage}--\pageref{lastpage}}
\maketitle

% Abstract of the paper
\begin{abstract}
We study the connection between galaxy morphology and host dark matter (DM) halo structure using cosmological simulations.
Introducing a new kinematic decomposition scheme, we robustly separate thin and thick discs and measure halo properties, including cosmic web locations, internal structures, and assembly histories. 
In the TNG50 simulation, we find that the orbital-circularity threshold for disc differentiation varies systematically with galaxy mass and redshift. 
Similarly, the energy threshold between stellar halos and inner galaxies depends on mass and redshift, minimizing at sub-Galactic halo mass where the circularity threshold approaches its peak. 
Revisiting galaxy size predictors, we show that disc sizes in TNG50 correlate with three structural parameters beyond virial mass and redshift: 1) a positive correlation with halo spin $\lambda$ across redshifts -- stronger than previously reported for zoom-in simulations but still weaker than the simple $r_{1/2}/R_{\rm vir} \propto \lambda$ scaling; 2) an anti-correlation with DM concentration $c$; 3)  larger discs in more actively accreting haloes. 
Disc mass fraction is higher in rounder haloes and in cosmic knots and filaments, implying that disc development needs both stable halo conditions and continuous material supply.  
Our methodology is public and adaptable to other simulations.
\end{abstract}

% Select between one and six entries from the list of approved keywords.
% Don't make up new ones.

%%%%%%%%%%%%%%%%%%%%%%%%%%%%%%%%%%%%%%%%%%%%%%%%%%

%%%%%%%%%%%%%%%%% BODY OF PAPER %%%%%%%%%%%%%%%%%%

\section{Introduction}\label{sec:intro}

In the standard paradigm of cosmic structure formation, dark-matter (DM) haloes provide the sites for galaxy formation.
Galaxies show dramatic structural diversity across redshifts. 
At certain mass scales, galaxies span two orders of magnitude in half-mass radius and surface brightness \citep{Sales22}, or exhibit drastically different morphologies ranging from large bulgeless disks \citep{Wang24} to compact red nuggets \citep{Damjanov11}.
It is natural to speculate that the morphologies of galaxies contain information of host DM haloes, and that the structural diversity of the haloes is correlated with that of the inhabitant galaxies.
The halo virial mass determines many galactic properties statistically, not only the stellar mass \citep[e.g.,][]{Yang03,Yang08,Yang12,Behroozi13,Moster13}, but also various aspects of galaxy morphology. 
Notably, the size of a galaxy is approximately $\sim 2\%$ of the virial radius \citep[e.g.,][]{Kravtsov13, Somerville18} (which is equivalent to halo mass at a given epoch); and galactic angular momentum (AM) has also been shown in cosmological hydrodynamical simulations to transit from a fast-flipping state to a more stable status after the halo reaches a characteristic mass scale, above which both stellar feedback and disc instability weaken \citep{Dekel20, Hopkins23}. 
Besides halo mass, which secondary properties of the DM halo are most relevant in regulating galactic morphology has become a pivotal question in understanding galaxy-halo connection and galaxy evolution \citep[e.g.,][]{Hearin13, Behroozi19, Chen21} that all semi-analytical and semi-empirical frameworks of galaxy evolution need to address. 

For example, semi-analytical models (SAMs) assign disc sizes to star-forming galaxies according to certain dark-matter halo properties. 
While which halo property is most relevant for disc size is still a open question \citep[e.g.,][]{Jiang19, Behroozi19, Shen24}, the classical prescription based on AM conservation is that disc size is proportional to the spin parameter $\lambda$ of the host halo \citep{Fall80, Mo98}, which, in turn, can be predicted by the tidal torque theory \citep{White84} or obtained from cosmological $N$-body simulations. 
This motivates some SAMs to predict the sizes of star-forming discs as $\rhalf \sim \lambda \Rv$, where $\Rv$ is the virial size of the host halo and $\lambda$ the instantaneous halo spin \citep[e.g.,][]{Somerville08}.
Other SAMs compute disc size using the disc AM accumulated over time from that of the circum-galactic medium, which in turn depends on some time average of the halo spin \citep[e.g.,][]{Benson12}.
Such recipes, with the cosmological average spin of DM haloes being $\sim 0.03$, agrees with the observed proportionality between galaxy sizes and halo sizes as inferred from abundance matching \citep{Kravtsov13}.
Galactic bulges form as major mergers of disc galaxies occur and bulge sizes reflect energy conservation during the mergers.  
Thus in almost all SAMs, disc galaxies are prevalent before bulges emerge. 
As such, the explanation of the morphological diversity of star-forming (dwarf) galaxies with SAMs is basically that, the most compact galaxies and the most diffuse galaxies populate DM haloes of the lowest and highest spin, respectively, by construction. 
However, some cosmological hydro simulations reveal a largely orthogonal story compared to the standard SAM view, in the sense that, in simulations, disc morphology is rare and unstable at high redshifts when cosmic accretion is intense, and when star formation and stellar feedback is bursty. 
It is after the formation of a central compact stellar bulge (or other forms of compact mass distribution) that long-lived, extended discs start to develop \citep{Dekel20, Hafen22, Yu23, Hopkins23}.
%\textcolor{red}{This can also be explained by recent model (Mo et al. 2023).}
Accordingly, the morphological diversity of galaxies in these simulations are not tied to the AM content of the haloes, but may be related to other structural parameters \citep[e.g.,][]{Jiang19}.

Galactic morphological diversity is not merely defined in terms of disc size, but also in terms of the properties of different morphological components, including but not limited to bulges, stellar haloes, thin discs, and thick discs. These morphological components occupy different regions in kinematic spaces, for example, stars in discs exhibit ordered rotation, whereas those in bulges and haloes move on random orbits \citep[e.g.,][]{Du19}. Investigations of galaxy morphology is however hindered by the difficulty in cleanly separating different morphological components.
Even with cosmological simulations, where kinematics information of stars is fully recorded, it is still not clear how to define the aforementioned components using self-consistent physical criteria. 

For example, the most widely used kinematic decomposition method would designate the stars with an orbital circularity above a fixed arbitrary threshold value as thin-disc stars \citep[e.g.,][]{Tacchella19, Sotillo-Ramos23,Yu21,Yu23}.
Similar arbitrary thresholds exist for thick discs and spheroidal components. 

Some of the recent developments in decomposition techniques start to address this issue by exploring variable thresholds in kinematics spaces \citep[e.g.,][]{Zana22} or by introducing methods that do not explicitly rely on any threshold \citep[e.g.,][]{Sokolowska17,Obreja18,Jagvaral22}.
However, in many studies that still rely on kinematics thresholds, different authors use different threshold values, or have different rules for assigning stars to different components, hindering efficient comparisons and reproducibility. 
Hence, to investigate the connections between galaxy morphology and DM halo structures, and to facilitate meaningful comparisons amongst different simulations, we first need to clarify morphological definitions and minimize arbitrariness in them.

The structural parameters of DM haloes include not merely the AM structure and the density-profile shape. 
They at least also include the 3D triaxial shape \citep{JingSuto02}, and broadly speaking, the mass assembly histories as well as the cosmological large-scale environments \citep[e.g.,][]{Wang07,Wang11,Wang18,Shi15}. 
Haloes acquire their AM from large-scale tidal torques \citep{White84}, and gain broken power-law density profiles from fast accretion in the early Universe followed by slow build-up of the outskirt at later times \citep[e.g.,][]{Zhao03, Zhao09, Ludlow13}. 
The 3D shapes also contain information of the cosmic web at large \citep{Forero-Romero14,Ganeshaiah18}.
Hence, the instantaneous structure of a DM halo are correlated with both its assembly history and the cosmic environment. 
It is natural to expect that galaxy morphology also contains information of these spatial and temporal factors \citep[see e.g.,][]{Mo23}. 

In this series of studies, we work towards a comprehensive exploration of the relationship between galactic morphology and the broadly-defined DM-halo structural properties. 
In particular, we aim at sorting out the answers to the following questions:
\begin{itemize}[leftmargin=*]
\item[1.] Can we simplify kinematic morphological decomposition of simulated galaxies into a robust and non-arbitrary workflow?  
\item[2.] Which DM-halo parameters are most tightly correlated with a chosen galaxy morphological parameter? For example, what halo properties constitute the best predictor for disc size? 
\item[3.] To what extent can we use DM halo properties alone to predict the morphological properties of the inhabitant galaxies? For example, from the point of view of semi-analytical or semi-empirical modeling, is it feasible at all to paint galaxies to haloes using a collection of secondary halo parameters, especially given that galaxies are shaped also by their own internal baryonic processes and that the DM may be redistributed by these processes \citep[e.g.,][]{Blumenthal86, Gnedin04, Pontzen13}?
\end{itemize}
This work (Paper I) addresses the first question and also touches base on the second question. 
In paper II, we address morphology-halo relations in more detail, with the help of machine-learning algorithms. 

Paper I is organized as follows. 
In \se{measurements}, we describe the morphological and structural quantities that we measure for the simulated galaxies.
In \se{decomposition}, we introduce our new kinematic decomposition scheme and comment on some insights that are immediately clear with this method.
We present a few correlations between galaxy morphological parameters and halo structural parameters in \se{result}, with an emphasis on which halo properties affect disc size. 
Different decomposition methods are compared in \se{discussion}, and we summarize this study in \se{conclusion}. 
%We emphasize that while this work uses a specific simulation suite, our methodology laid out here is easily adaptable to other cosmological simulations...
Throughout this study, we focus on central galaxies unless otherwise stated, and define haloes as spherical overdensities that are 200 times the critical density of the Universe, and assume cosmological parameters adopted in the simulations analyzed.

%--------------------------------------------------
%--------------------------------------------------

\section{Simulation Catalog and Measurements}\label{sec:measurements}

%The galaxies and host dark matter haloes are from IllustrisTNG project \citep{Springel18,Nelson18,Marinacci18,Pillepich18a,Naiman18}, which is a suite of cosmological, gravo-magneto-hydrodynamical simulations of galaxy formation and evolution in comoving boxes of side 50, 100 and 300 cMpc. Each run has been conducted with moving-mesh hydrodynamical code AREPO \citep{Springel10} and assumes $\Lambda$CDM cosmology with parameters from \citep{Planck2016}, i.e. $\Omega_{\rm M,0}$ = 0.3089, $\Omega_{\Lambda,0}$ = 0.6911, $\Omega_{\rm b,0}$ = 0.0486, $H_{0}$ = 67.74 km s$^{-1}$ Mpc$^{-1}$. Physical galaxy formation model and its numerical details including star formation, stellar evolution, chemical enrichment, gas cooling, galactic outflows, stellar and black hole feedback are described in \citep{Pillepich18b,Weinberger17}. haloes and subhaloes are identified by Friends-of-Friends \citep[FoF,][]{Davis85} and SUBFIND \citep{Springel01} algorithms

To extract statistical galaxy-halo connections that involve detailed morphological information from cosmological hydro-simulations, we need a simulation sample with both cosmologically meaningful box size and adequate numerical resolution. 
In this work, we use the highest resolution run of the Illustris-TNG suite,  TNG50 \citep{Pillepich19,Nelson19}, which has a gas particle mass of $10^{4.93}\Msun$, a DM particle mass of  $10^{5.65}\Msun$, and a gravitational softening length for collisionless (gas) particles of 0.288 (0.074) comoving kpc. 
We limit our analysis to the galaxies with a half-stellar-mass radius at least two times the collisionless softening length, and those with more than 1000 stellar particles as well as 1000 DM particles within virial radius.
We use the galaxies and haloes in the public catalogs\footnote{\href{https://www.tng-project.org/data/downloads/TNG50-1}{https://www.tng-project.org/data/downloads/TNG50-1/}}, for which haloes are identified with the Friends-of-Friends \citep[FoF,][]{Davis85} and {\tt SUBFIND} \citep{Springel01} algorithms, and are linked accross snapshots with the {\tt SUBLINK} merger tree algorithm \citep{Rodriguez-Gomez15}.
For certain analysis, we use DM haloes in the DM-only simulation of the same initial conditions and focus on the haloes that have a matched counterpart in the hydro-simulation according to the public {\tt Subhalo Matching To Dark} catalog \footnote{\href{https://www.tng-project.org/data/docs/specifications/\#sec5d}{https://www.tng-project.org/data/docs/specifications/\#sec5d}}. 
To reveal potential redshift trends, we analyze the simulations at $z=0$, 1, 2, and 4.
We use the halo catalogs mostly only for obtaining the global properties such as halo mass and galaxy mass. 
For the structural parameters, we perform measurements using the particle data and our in-house developed programs that are highly modular, so that such measurements can be easily extended to other simulations in the future for  comparisons. 
We focus on central galaxies, those with center coordinates instantaneously not within the virial radius of any other halo. 
We exclude the systems that are perturbed by ongoing mergers, which hinder accurate structural measurements.
We make our measurements as well as our particle-data reduction pipeline publicly available at 
\href{https://github.com/JinningLianggithub/MorphDecom}{https://github.com/JinningLianggithub/MorphDecom}.
Below, we list all the structural parameters considered in this study and describe their operational definitions, summarized in \Tab{quantities}.

\begin{table*}
\centering
\begin{tabular}{lllllllllll}
\hline
\hline
%\multicolumn{11}{c}{Quantities used in this work} \\ \hline
                                                                
\multicolumn{2}{l}{Halo quantities} & \multicolumn{7}{l}{Description} & \multicolumn{2}{l}{Reference} \\ \hline

\multicolumn{2}{l}{Halo concentration $c$}        & \multicolumn{7}{P{7cm}}{$c=\Rv/r_{-2}$ based on fitting Einasto profile}           & \multicolumn{2}{l}{\cite{Einasto65}}         \\
\multicolumn{2}{l}{Halo spin $\lambda$}        & \multicolumn{7}{l}{$\jv/(\sqrt{2}\Rv\Vv)$}           & \multicolumn{2}{l}{\cite{Bullock01}}         \\
\multicolumn{2}{l}{Late-time accretion parameter $\beta$}        & \multicolumn{7}{l}{as in $\Mv(z)\propto (1+z)^\beta e^{-\gamma z}$, MAH slope at late times}           & \multicolumn{2}{l}{\cite{McBride09}}         \\
\multicolumn{2}{l}{Early-time accretion parameter $\gamma$}        & \multicolumn{7}{l}{as in $\Mv(z)\propto (1+z)^\beta e^{-\gamma z}$, MAH slope at early times}           & \multicolumn{2}{l}{\cite{McBride09}}         \\
\multicolumn{2}{l}{Halo axis ratio $q$}        & \multicolumn{7}{P{7cm}}{ratio between intermediate axis and major axis $b/a$ calculated from eigenvalues of shape tensors}           & \multicolumn{2}{l}{\cite{Allgood06}}         \\ 
\multicolumn{2}{l}{Environment}        & \multicolumn{7}{P{7cm}}{location in cosmic web as {\it knot}, {\it filament}, {\it sheet} or {\it void}, based on local eigenvalues of tidal shear tensor}           & \multicolumn{2}{l}{\cite{Hahn07}}         \\ \hline

\multicolumn{2}{l}{Galaxy (stellar-particle) quantities} & \multicolumn{7}{l}{Description} & \multicolumn{2}{l}{Reference} \\ \hline

\multicolumn{2}{l}{Size $\rhalf$}        & \multicolumn{7}{P{7cm}}{3D radius within which the enclosed stellar mass is equal to half of total stellar mass within the halo}           & \multicolumn{2}{l}{public TNG50 catalog}         \\
\multicolumn{2}{l}{Normalized specific binding energy $e_{\rm b}$}        & \multicolumn{7}{P{7cm}}{Specific binding energy, $E$, scaled by the absolute value of $E$ of the most bound particle, $|E|_{\rm max}$}           & \multicolumn{2}{l}{\cite{Domenech-Moral12}}         \\
\multicolumn{2}{l}{Circularity $\eta$}        & \multicolumn{7}{P{7cm}}{Azimuth specific AM $\jz$ scaled by the maximal AM with the same binding energy $\jc(E)$}           & \multicolumn{2}{l}{\cite{Abadi03}}         \\
\multicolumn{2}{l}{Polarity $\epsilon$}        & \multicolumn{7}{P{7cm}}{Non-azimuth specific AM $\jp$ scaled by the maximal AM with the same binding energy $\jc(E)$}           & \multicolumn{2}{l}{\cite{Domenech-Moral12}}         \\
\multicolumn{2}{l}{Circularity threshold $\eta_{\rm cut}$}        & \multicolumn{7}{P{7cm}}{Intersection of the $\eta$ distributions of the two discy components found by the Gaussian Mixture Models algorithm -- for separating thin and thick discs}           & \multicolumn{2}{l}{This work}         \\
\multicolumn{2}{l}{Energy threshold $e_{\rm cut}$}        & \multicolumn{7}{P{7cm}}{Local minimum in the $e_{\rm b}$ distribution -- for separating high-energy and low-energy stellar components}           & \multicolumn{2}{l}{\cite{Zana22}}         \\ 
\multicolumn{2}{l}{Mass fraction $f_X$}        & \multicolumn{7}{P{7cm}}{$f_{\rm X}=M_{\star,X}/\Ms$, where $M_{\star,X}$ is the mass of morphological component $X$}           & \multicolumn{2}{l}{ }         \\ \hline
\hline
\end{tabular}
\caption{Structural parameters of DM haloes and morphological quantities of galaxies (as well as kinematics quantities of stellar particles) used in this study.}\label{tab:quantities}
\end{table*}

%--------------------------------------------------

\subsection{Dark-Matter Halo Quantities}\label{sec:HaloQuantities}

\begin{itemize}[leftmargin=*]

\item \textbf{Halo mass and virial radius} $\Mv$ is defined as the enclosed {\it total} mass within the sphere of radius $\Rv$, the radius within which the average total density is $\Delta$ times the critical density of the Universe, with $\Delta=200$, consistent with the definition used for the {\tt SUBFIND} catalog. 

\item \textbf{The circular velocity} is defined as $V_{\rm circ,dm}=\sqrt{GM_{\rm dm}(<r)/r}$, where $M_{\rm dm}(<r)$ is the enclosed DM mass within radius $r$.
The maximum circular velocity, $\Vmax$, is the maximum value of the $V_{\rm circ,dm}(r)$, and $\rmax$ is the radius at which $V_{\rm circ,dm}=\Vmax$. 
Note that the definitions for $\Vmax$ and $\rmax$ are different from those in the public {\tt SUBFIND} catalog, the latter of which are based on the total mass profiles. 

\item \textbf{Halo concentration} describes the compactness of the DM halo and reflects the depth of the gravitational potential well at fixed halo mass, defined as $c=\Rv/r_{-2}$, where $r_{-2}$ is scale radius at which the logarithmic density slope is $-2$.
To measure concentration, the most commonly used approach is to perform parametric fits to the density profile, assuming, e.g., the \citet[][hereafter NFW]{NFW97} functional form.
Haloes are more accurately described by the \citet{Einasto65} profile, 
\be
\rho_{\rm Einasto}=\rho_{-2}\exp\left\{-2n\left[\left(\frac{r}{r_{-2}}\right)^{1/n}-1\right]\right\},
\ee
at the expense of just one additional parameter compared to NFW. 
This additional degree of freedom is important for capturing baryonic and environmental impacts on halo structure.
Here, $n$ is the Einasto shape index, $\rho_{-2}$ is the density at $r_{-2}$, linked to the virial mass via $\rho_{-2} = \Mv / (4\pi h^3 n \gamma(3n,x(\Rv))$, with $h\equiv r_{-2}/(2n)^n$, $\gamma(a,x)$ the non-normalized incomplete gamma function, and $x\equiv 2n(r/r_{-2})^{1/n}$.  
We have experimented with different measurements of halo concentration, including parametric fits of NFW and Einasto functional forms, and non-parametric proxies constructed using $\Vmax$ and $\rmax$, $\tilde{V}_{\rm max}$, as proposed by \citet{Bose19}, as well as a straightforward $c=\Rv/r_{-2}$, with the scale radius $r_{-2}$ measured from the density profile. 
%The details of the measurements are provided in Appendix XXX... 
We find that the concentration based on fitting the Einasto circular velocity profile carries the least fitting error and yields the strongest correlations with galaxy morphologies, and thus use it as the default definition for concentration.
We exclude 10\% of the haloes with the largest reduced $\chi^2$ in the profile fitting, as these are likely perturbed by ongoing mergers.  

\item \textbf{The halo spin parameter} is defined as $\lambda = \jv /(\sqrt{2}\Rv\Vv)$,
following \citet{Bullock01}, where $\jv$ is the specific AM within the virial radius, and $\Vv$ is the circular velocity at virial radius. 

\item \textbf{The mass assembly history} (MAH) of a DM halo is defined as the mass of the main branch progenitor as a function of redshift, $\Mv(z)$. 
The MAHs from the simulations are arrays of length up to the total number of snapshots, and are thus not straightforward to use in the search for correlations between galaxy morphology and halo MAH.  
To reduce the dimension of MAHs, we fit the MAHs using the functional form introduced by \citet{McBride09},
\be
\Mv(z)=\Mv (1+z)^\beta e^{-\gamma z},
\ee
where $\beta$ and $\gamma$ are the two free parameters describing the speed of halo growth at late and early times, respectively. 
Here the exponential growth is an analytic prediction in the EdS regime \citep{Dekel13}, and the later power-law redshift dependence is an empirical results based on simulations. 
We use $\beta-\gamma$ as a proxy for the recent accretion rate, because the accretion rate
\be
\frac{\dd \ln{\Mv(z)}}{\dd z}=\frac{\beta}{1+z}-\gamma
\ee
is approximately $\beta-\gamma$ at $z\sim0$. 
At higher redshift, it can be also shown that the average accretion rate within a dynamical time is strongly correlated with $\beta-\gamma$. 
As already discussed in \citet{McBride09}, MAHs can be classified according to the value of $\beta-\gamma$. 
The more actively accreting a halo is, the more negative $\beta-\gamma$ is. 
For haloes that are halted in mass growth or have started losing mass, $\beta-\gamma\ga 0$. 
Hence, while we focus on central haloes, we regard $\beta-\gamma>0$ as an implication of backsplash or at least an indicator of environmental influence.

\begin{figure*}	
\includegraphics[width=\textwidth]{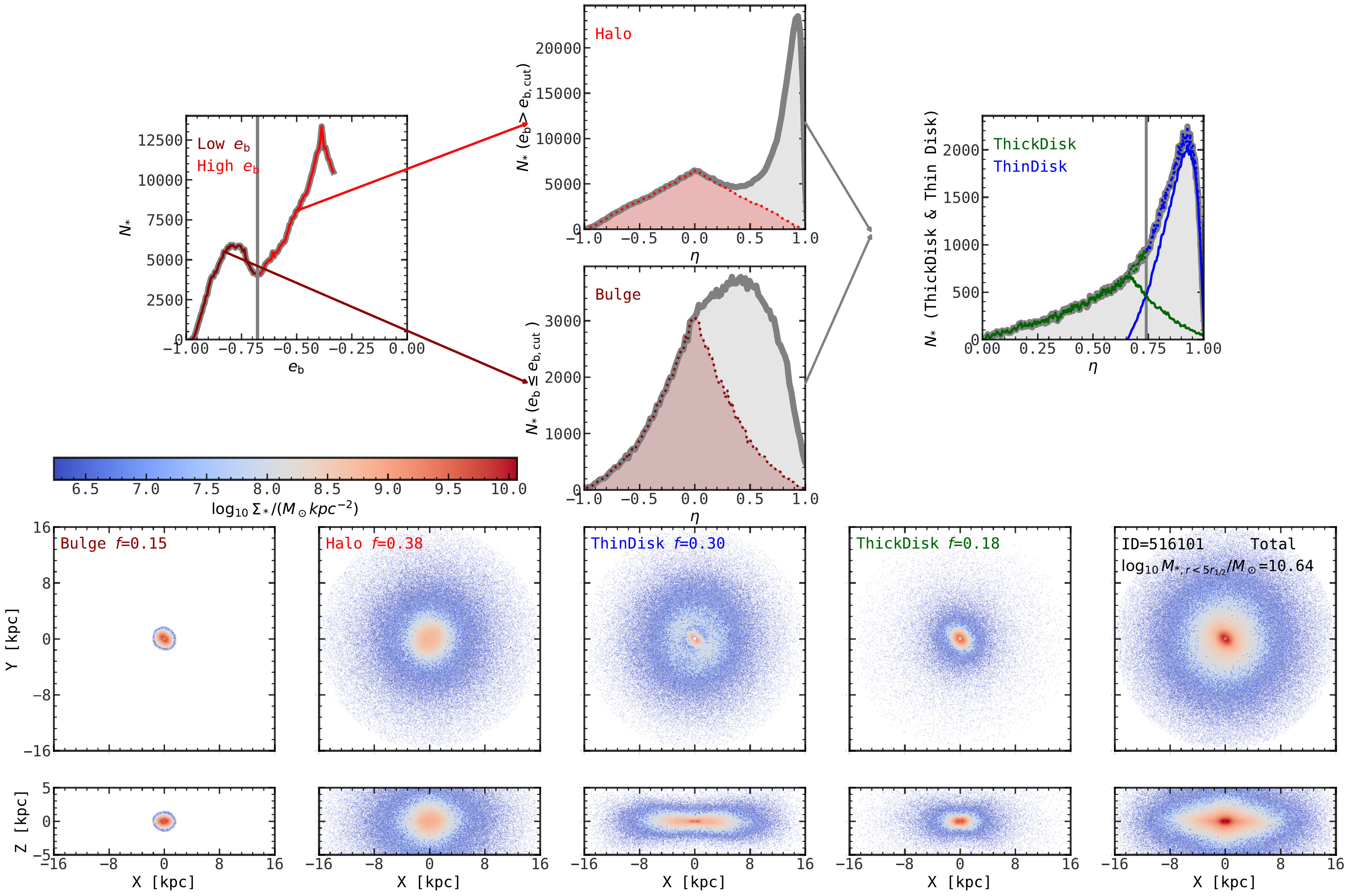}
    \caption{Illustration of our new decomposition method, with a Milky-Way-mass galaxy at $z$=0. 
    {\it Upper panels}: the distributions of specific binding energy $\eb$ and circularity $\eta$. 
    In the left panel, the thick gray curve stands for the $\eb$ distribution of the whole galaxy, with the brighter red curve marking the high-energy component and darker red curve standing for the low-energy component, separated by the vertical gray line indicating the threshold $\ecut$. 
    In the upper middle column, we show the circularity distributions of the low-energy and high-energy components, respectively -- the bulge and stellar halo correspond to the symmetric distributions highlighted in red, where the positive $\eta$ parts are obtained from Monte Carlo sampling. 
    For the remaining stellar particles, as indicated by the gray shaded $\eta>0$ areas in the middle columns and combined in the right panel, they are first classified into two subgroups with the GMM algorithm, and are further split into thin and thick discs according to the threshold $\etacut$. The green and blue lines are the smoothed Gaussian components provided by GMM, whose intersection is used as the threshold $\etacut$.
    {\it Lower panels}: face-on and edge-on views of the resulting bulge, halo, thin disc, and thick disc components, coloured by surface density, with the galaxy ID and its stellar mass within 5 half-stellar-mass radius quoted.}
    \label{fig:MW-method}
\end{figure*}

\item \textbf{The 3D shape} of DM haloes can be characterized by the eigen-values of the inertia tensor \citep{Allgood06},
\be
\mathcal{S}=\frac{1}{M}\sum_{k}m_k\bm{r}_{k,i}\bm{r}_{k,j},
\ee
where the summation is over all the DM particles within the ellipsoid of interest, $\bm{r}_{k,i}$ is the component of the position vector of the $k$th particle along axis $i$, and $M=\Sigma_{k}m_k$ is the total mass within the ellipsoid. 
The eigenvalues of $\mathcal{S}$ are proportional to the squares of the semi-axes ($a \geq b \geq c$) of the ellipsoid that describes the spatial distribution of the particles. 
We measure the eigenvalues iteratively, starting from the virial sphere until convergence, following the algorithm of \citet{Tomassetti16}. 
The shapes are expressed in terms of the 3D axis ratios, e.g., $q=b/a$ and $p=c/b$.

\begin{table*}
\centering
\begin{tabular}{ccccccccccccc}
\hline
         & \multicolumn{4}{c}{$F_V$}       & \multicolumn{4}{c}{$F_M$}       & \multicolumn{4}{c}{$N_{\rm cen}$}     \\
Redshift & Knot & Filament & Sheet & Void & Knot & Filament & Sheet & Void & Knot & Filament & Sheet & Void \\ \hline
0        & 0.005    & 0.137        & 0.367     & 0.491    & 0.350    & 0.374        & 0.191     & 0.085    & 150    & 1297        & 3136     & 506    \\
1        & 0.006    & 0.116        & 0.342     & 0.536    & 0.223    & 0.353        & 0.268     & 0.156    & 160    & 1603        & 3620     & 596    \\
2        & 0.006    & 0.094        & 0.311     & 0.588    & 0.128    & 0.304        & 0.322     & 0.246    & 166    & 1646        & 3464     & 650    \\
4        & 0.004    & 0.056        & 0.237     & 0.703    & 0.041    & 0.180        & 0.332     & 0.447    & 76    & 758        & 1768     & 427    \\ \hline
\end{tabular}
\caption{Volume fraction, mass fraction and number of TNG50 central galaxies residing in knots, filaments, sheets, and voids at different redshifts.}
\end{table*}

\item \textbf{The large-scale environment} of a DM halo can be characterized as the position within the cosmic web.
%A large number of methods have been proposed to classify cosmic web including adapted minimal spanning tree \citep[MST,][]{Alpaslan14}, Bisous \citep{Tempel14}, Filament Identification using NodEs \citep[FINE,][]{Gonzalez10}, Tidal shear tensor \citep[T-web][]{Hahn07,Forero-Romero09}, Velocity shear tensor \citep[V-web][]{Hoffman12}, Lagrangian perturbation theory \citep[CLASSIC][]{Kitaura12}, NEXUS \citep{Cautun13}, Multiscale Morphology Filter-2 \citep[MMF-2][]{Arag0n-Calvo07}, Spineweb \citep{Aragon-Calvo10}, DisPerSE \citep{Sousbie11}, ORIGAMI \citep{Falck12}, Multi-Stream Web Analysis \citep[MSWA][]{Ramachandra15}. Details about these methods can be seen in recent review \citep{Libeskind18}. 
We classify cosmic web using the eigenvalues of the deformation tensor \citep[e.g.,][]{Hahn07,Forero-Romero09},
$T_{i,j}$.
Based on the analysis of large-scale structures using the Zel'dolvich approximation, cosmic web can be classified by counting the number of eigenvalues of the deformation tensor exceeding a threshold $\lambdath$ (chosen to be 0.4 in this work, see below), such that 0, 1, 2, or 3 eigenvalues exceeding $\lambdath$ corresponds to {\it void}, {\it sheet}, {\it filament}, or {\it knot}, respectively. 
To determine the tidal tensor $T_{i,j}$ and the eigenvalues at each coordinate, we follow the steps laid out in \citet{Martizzi19}, adapting the public code complementary to the work of \citet{Yang22}\footnote{
 \href{https://github.com/WangYun1995/Cosmic-Web-Classification-using-the-HessianMatrix}{https://github.com/WangYun1995/Cosmic-Web-Classification-using-the-HessianMatrix}.}.
In particular, first, starting from the Poisson equation:
\be\label{eq:Poisson}
\nabla^2\phi(\bm{x})=4\pi G \bar{\rho}\delta(\bm{x}),
\ee

where $G$ is the gravitational constant, $\bar{\rho}(z)$ is the mean matter density of the Universe, and $\delta(\bm{x})\equiv \rho(\bm{x})/\bar{\rho}-1$ is the overdensity, we compute the overdensity field by interpolating the mass of each particle to a $512^3$ Cartesian grid using a cloud-in-cell method smoothed by a Gaussian filter of scale $R_{\rm G}=0.5$ckpc$h^{-1}$. 
Second, to facilitate the computation of the tidal tensor, we express the gravitational potential in units of $4\pi G \bar{\rho}$ such that \eq{Poisson} becomes simply $\nabla^2\phi(\bm{x})=\delta(\bm{x})$.
With the Fourier transform of the overdensity field, $\delta_{\bm{k}}$, we can compute the Fourier transform of the tidal tensor as
\be
\Psi_{i,j,\bm{k}}=k_i k_j \phi_{\bm{k}},\ \ i,j=1,2,3,
\ee
where $k_i$ is wave number in Fourier space. 
By performing the inverse FFT of $\Psi_{i,j,\bm{k}}$ for each $(i,j)$, one can obtain the deformation tensor element $T_{i,j}(\bm{x})$ at each point in the original space. 
Finally, we solve
\be
\text{det}\left(\bm{T}(\bm{x}) - \lambda(\bm{x})\bm{I}\right)=0
\ee
for the eigenvalues $\lambda_1(\bm{x})>\lambda_2(\bm{x})>\lambda_3(\bm{x})$.
The parameter values of $\lambda_{\rm th}=0.4$ and $R_{G}=0.5$ cMpc$h^{-1}$ are chosen based on our visual comparison of the resulting cosmic-web classification and the density map of the simulation, and are in the same ballpark as the values reported in the literature \citep[e.g.,][]{Martizzi19}.

\end{itemize}

%--------------------------------------------------

\subsection{Galaxy Quantities}\label{sec:GalaxyQuantities}

\begin{itemize}[leftmargin=*]

\item \textbf{Half-stellar-mass radius} $\rhalf$ is the 3D radius within which the enclosed stellar mass is equal to half of total stellar mass within the halo, provided by the public TNG50 catalog. 

\item \textbf{Stellar mass} $\Ms$ -- instead of considering all the stellar particles within the virial radius, we define stellar mass as the stellar mass within 5$\rhalf$. \footnote{Where to crop the galaxy is somewhat arbitrary in the literature, our choice is comparable to the widely used choice of cropping at 0.1-0.2 $\Rv$ \citep[e.g.,][]{Obreja18} because $\rhalf$ is believed to be comparable to 0.02$\Rv$ \citep[e.g.,][]{Kravtsov13}. We stick to multiples of the stellar half-mass radii because our focus is on galaxy-halo connection and thus we do not want to introduce a halo scale in galactic measurements.}
We have verified that this can exclude most of the merger relics of satellite galaxies but at the same time retaining most of the smooth stellar halo of the central galaxies. 
We exclude the wind particles in TNG50 from the stellar mass.

\end{itemize}

%\item \textbf{Parameters for kinematic morphological decomposition -- circularity $\eta$, polarity $\epsilon$, and normalized binding energy $\eb$}

We base our morphological decomposition on the energy structure and AM structure of the stellar particles within 5$\rhalf$.
We present the workflow of our method in \se{decomposition}, but define the relevant quantities here. 
For each stellar particle, we consider two AM related parameters, the {\it circularity} $\eta$ and the {\it polarity} $\epsilon$, as well as one energy parameter, $\eb$, defined as follows. 

\begin{itemize}[leftmargin=*]

\item \textbf{Circularity $\eta$} \citep[e.g.,][]{Abadi03} is the ratio of the azimuthal AM $\jz$ to the maximum AM $\jc(E)$, 

\be\label{eq:circularity}
\eta \equiv \frac{\jz}{\jc(E)},
\ee

where $\jz=\bm{j} \cdot \hat{\bm z}$, with $\hat{\bm z}$ the unit vector of the total AM of the galaxy (calculated with all the stellar particles within 5$\rhalf$), and $\jc(E)$ is the AM of a circular orbit of the same specific orbital energy $E$ as the particle of interest has.
$\jc(E)$ is obtained by following its definition, $j_{\rm circ}(E)=\rc(E)\Vc(E)$, where $\Vc(E)$ is the circular velocity at given specific orbital energy $E$, and $\rc(E)$ is the radius of this circular orbit. 
The details of $\jc(E)$ evaluation are given in \se{decomposition}. Circularity measures the extent to which a particle belongs to the galactic disc: $\eta=1$ means maximal degree of coherent rotation and the particle belongs to the thin disk; $\eta\sim0$ refers to random motion; and $\eta=-1$ means counter-rotation.

\item \textbf{Polarity $\epsilon$} is the counterpart of circularity for the polar component of the AM, $\bm{j_{\rm p}}=\bm{j}-\bm{j_{\rm z}}$, is denoted by $\epsilon\equiv \jp/\jc(E)$, which we dub the {\it polarity} parameter \citep[e.g.,][]{Domenech-Moral12}.  
This parameter will be useful, together with the energy parameter $\eb$, when we split galactic discs further into thin and thick components.

\item \textbf{The normalized binding energy $\eb$} measures how bound a stellar particle is, and is defined as $e_{\rm b}=E/|E|_{\rm max}$, where $|E|_{\rm max}$ is the absolute value of the specific energy of the most bound particle.
Via detecting a local minimum in the distribution of $\eb$, we can decompose a galaxy into a higher-energy component and a lower-energy component.

\item \textbf{The mass fraction of a morphological component} is given by $f_{\rm X}=M_{\star,X}/\Ms$ where $M_{\star,X}$ is the mass of morphological component $X$ (bulge, stellar halo, thin disc, or thick disc), the identification of which is detailed in \se{decomposition}. 
$f_{\rm Disk}$ refers to the total disk fraction where the thin and thick components are combined.
Our method currently does not support bar identification.
%Note that some galaxies might not have all the four components, so $f_{\rm X}$ can be 0.

\end{itemize}

%\subsection{Decomposition Method Overview}\label{sec:DecompositionOverview}
%With the developments of high resolution cosmological simulation, it is possible to apply different algorithms to decompose galaxies into distinct structures including halo, bulge, disky bulge (pseudo bulge), thin disc (cold disc), thick disc (warm disc) and bar. Recent methodologies have been proposed by \cite{Obreja18}; \cite{Du19}; \cite{Du20}; \cite{Proctor23}, which based on Gaussian Mixture Model (GMM) using structural and kinematical information. Except applying machine learning algorithm, many work including \cite{Zana22} and \cite{Zhu22} also focus on setting threshold for structural or kinematical parameters and splitting stellar particles by thresholds. While the latter method is more physical compared to machine learning algorithms, the setting for threshold is arbitrary. 

%--------------------------------------------------
%--------------------------------------------------
\section{A New Morphological Decomposition Procedure}\label{sec:decomposition}

Most previous studies on kinematic morphological decomposition adopt constant thresholds in circularity \citep[e.g.,][]{Yu21, Yu23, Tacchella19, Sotillo-Ramos23}.
For example, a significant number of studies assign stars with $\eta\ga 0.7$ to disc \citep[e.g.,][]{Aumer13,Sokolowska17}. 
Most works do not consider the energy structure of stellar mass.

\citet{Zana22} improves this simple scheme by considering a local minimum in the energy distribution to split star particles into a high-energy population and a dynamically colder population.
Since the stellar halo is the most extended part of a galaxy, and stellar bulge is the dense part in the central region of a galaxy, it is natural to define stellar halo and bulge, respectively, in the higher-energy and colder components, respectively.
Then they still use a constant circularity threshold to further identify the discs.  

That is, a physically motivated criterion is used in the energy space, but the circularity criteria is still constant.
The difficulty for a self-consistent criteria in the AM space is simply that there is usually no clear local minimum in the circularity distribution. 
However, in the 3D parameter space spanned by $\eta$, $\epsilon$, and $\eb$, star particles can be more clearly separated into different groups.
\citet{Du19} has demonstrated this using a unsupervised machine-learning approach: they first split stars into several groups in the 3D space, before assigning them into different morphological components according to whether the median circularities and median energies of the groups fall into the $\eta$ and $\eb$ regimes that define the morphological components.
But again, constant and arbitrary circularity thresholds (as well as energy thresholds) are used. 

Incorporating the wisdom in \citet{Zana22} and \citet{Du19}, here we introduce a new decomposition scheme, minimizing the arbitrariness in these methods. 
Basically, we use the Gaussian-Mixture-Models algorithm (GMM, as implemented in 
the Python package {\tt scikit-learn}) to separate thin and thick discs, after applying a \citeauthor{Zana22} style energy cut and identifying the bulge and stellar halo.
We enforce GMM to search for two groups in the $\eta$-$\epsilon$-$\eb$ space after removing the random-motion supported bulge and halo, and find that GMM can robustly return us a higher-$\eta$ group, which corresponds to the thin disc, and a lower-circularity (and usually also higher-energy) subsystem, which corresponds to the thick disc. 
The detailed procedure is as follows.

\begin{figure*}	
\includegraphics[width=0.8\textwidth]{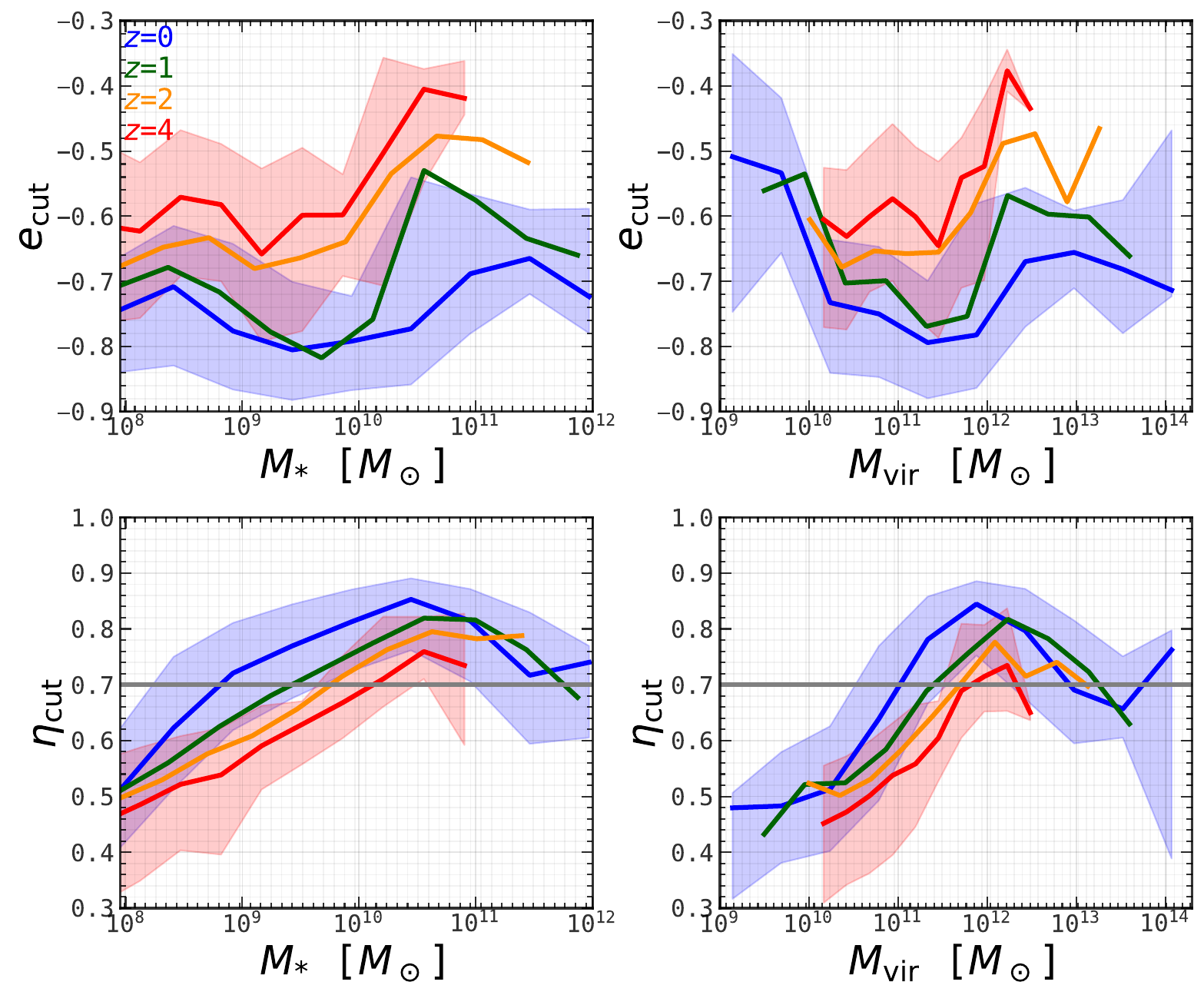}
    \caption{The energy threshold $\ecut$ (upper) and orbital-circularity threshold $\etacut$ (lower) for morphological decomposition, as a function of galaxy stellar mass (left) or DM halo mass (right), at different redshifts, as indicated. 
    The blue and red bands indicate the 16th and 84th percentiles for redshift 0 and 4, respectively. 
    The variances at given mass are similar for other two redshifts and are thus not shown to avoid crowdedness. 
    The horizontal gray lines in the lower panels represent the widely-used constant circularity threshold for separating thin and thick disc, $\etacut=0.7$. 
    Both thresholds vary systematically with mass and redshift, with peak $\etacut$ and minimal $\ecut$ occuring at the characteristic halo mass scale of $\Mv\sim10^{11.5-12}\Msun$, insensitive to redshift. The trend is more pronounced with halo mass than with stellar mass.
  }
    \label{fig:EcutEtacut}
\end{figure*}

\begin{itemize}[leftmargin=*]
\item[1.] Use the subhalo position from the TNG50 subhalo catalog, which is the location of the particle of the minimum gravitational potential, as galaxy centre. 
\item[2.] In order to compute energy and AM for kinematic decomposition, one needs to know the specific potential energy of each particle. 
This information is not readily available in the TNG particle data, so we re-evaluate the potential of a galaxy using all its particles using the KDTree algorithm {\tt pytreegrav}\footnote{\href{https://github.com/mikegrudic/pytreegrav}{https://github.com/mikegrudic/pytreegrav}}. 
We normalize the potential by setting that of the farthest particle to zero. This is effectively equivalent to setting the potential to zero at the virial radius. 
\item[3.] Compute the AM of the galaxy from the stellar particles within 5$\rhalf$, and rotate the galaxy such that the $z$ axis is aligned with the AM vector. 
\item[4.] Set 100 logarithmically spaced bins of the cylindrical radius $R=\sqrt{x^2+y^2}$, between the galaxy centre and the farthest particle. 
For each bin, we compute the gravitational potential $\Phic$ and the gravitational acceleration $\bm{a}_{\rm circ}$ using the KDTree from Step 2 at four positions, $(x,y,z)=$ $(R,0,0)$, $(0,R,0)$, $(-R,0,0)$, $(0,-R,0)$ and take the average values of the four measurements as the values for $\Phic$ and $\bm{a}_{\rm circ}$. 
With the gravitational acceleration in the galactic plane, we calculate the circular velocity, $\Vc=\sqrt{|\bm{a}_{\rm circ}\times\bm{r}_{\rm circ}|}$, the specific circular AM, $\jc = R\Vc$, and the specific binding energy in the galactic plane, $\Ec =\Vc^2/2+\Phic$. 
Thus we obtain a profile of $\jc(\Ec)$, which we turn into an interpolation function and use it to get the $\jc$ of individual particles given their $\Ec$.
\item[5.] For each stellar particle, we calculate the normalized specific binding energy $e_{\rm b}$, circularity $\eta$, and polarity $\epsilon$, per their definitions (\se{GalaxyQuantities}).
\item[6.] Remove unbound ($\eb\geq 0$) particles and rare particles of extreme values of circularity and polarity of $|\eta|\geq 1.5$ and $|\epsilon|\geq 1.5$.
\item[7.] Look for a local minimum in the $\eb$ distribution, register the corresponding $\eb$ value as $\ecut$, the energy threshold for separating stellar halo from the bulge. 
Here, we mostly follow the method of \citeauthor{Zana22}, but revise it to better accommodate dwarf galaxies with small numbers of stellar particles. 
In \app{EnergyThreshold}, we recap their method and detail our improvements. 
\item[8.] Split stellar particles into a more bound component, $\eb \leq \ecut$, and a less bound component, $\eb> \ecut$.
The stellar halo and bulge are identified as the random-motion supported subsets of the less bound and the more bound components, respectively. 
Since these spheroidal components are assumed to have zero net rotation, we identify them as the particle distributions symmetric about $\eta = 0$.
In particular, we identify the negative-circularity part and its mirrored distribution about $\eta=0$, and draw the $\eta>0$ particles to be assigned to the spheroidal components from the mirrored distributions via Monte Carlo sampling.
\item[9.] With the spheroids determined, we collect the remaining stellar particles altogether regardless of their binding energy -- these particles all have positive circularity and thus belong to discs. \footnote{We neglect counter-rotating disks in this study.}
We identify a circularity threshold, $\etacut$, to split them into thick and thin discs (if they both exist). 
Naturally, similar to the search for a energy threshold, $\ecut$, the idea is to obtain a threshold in the circularity distribution. 
As stated, quite often there is no local minimum in the one-point function of circularity. 
Hence, we split the stars into two groups in the 3D phase space spanned by $\eta$, $\epsilon$, and $e_{\rm b}$ using GMM, by enforcing two Gaussian components. 
\app{GMM} provides further details on how the GMM algorithm operates and \app{3DGMM} shows an example of gaussian components in $\eta$-$\epsilon$-$e_{\rm b}$ space.
The GMM fit is performed 10 times with different initialization for robust classification \citep{Du19}. 
Next, we identify the intersection of the circularity distribution of the two Gaussian components as the threshold $\etacut$. 
The stellar particles with $\eta \geq \eta_{\rm cut}$ are then assigned to the thin disc and the rest are thick-disc particles. 
If no intersection point is detected, which can happen when one of the two Gaussian components is significantly subdominant, then all the positive-circularity particles are assigned to the thin disc.
\item[10.] Finally, we measure the density profiles of different morphological components for future work. 
The details will be presented in Paper II.
To ensure accurate measurements, we impose that each spheroidal (disky) component should have at least 30 (100) star particles. 
We redistribute particles between bulge and halo or between thin disc and thick disc if the particle of a certain component is below the minimum value, e.g., if a bulge has fewer than 30 particles, the bulge is absorbed into the stellar halo. 
If the numbers of star particles of the thin disc and the thick disc are both below 30, they will be re-assigned to bulge or halo based on whether their $\eb$ is below or above $\ecut$.
\end{itemize}

\Fig{MW-method} illustrates our morphological decomposition method with an example of a Milky-Way-mass disc galaxy ($\Mstar=10^{10.64}\Msun$). 
As one can see in the upper left panel, our procedure successfully identifies the most pronounced local minimum in the energy distribution at $\ecut\simeq-0.68$, and ignores smaller local minima. 
The middle column in the upper panel shows the stellar halo and the stellar bulge, as the two components with symmetric circularity distributions centred around $\eta\sim0$. 
The remaining stars that have positive circularity are combined, as shown in the upper right panel. 
In this representative case, there is no local minimum in the circularity distribution, while the GMM algorithm successfully identifies two distinct groups in the 3D $\eta$-$\epsilon$-$\eb$ space. 
The two groups have circularity distributions intersecting at $\eta\simeq 0.8$, which we adopt as the circularity threshold, $\etacut$.
%After finding the threshold, $\etacut$, we use it to decompose total distribution into thin disc and thick disc. 
In the lower panels of \Fig{MW-method}, we show the stellar surface density maps of the four components and the whole galaxy. 
The mass fractions of the bulge, the stellar halo, the thin disc, and the thick disc are 0.15, 0.38, 0.25, 0.22, respectively. 
In \app{vis} we show more examples for different kinds of systems, including rare extreme cases such as bulgeless discs and pure spheroids.
We have manually inspected many representative cases of the decomposition results based on the aforementioned procedure and find it to be quite robust.  

Using our methods, the kinematic-decomposition thresholds are no longer constants.
Interestingly, both $\etacut$ and $\eb$ systematically vary with galaxy mass, DM-halo mass, and redshift, as shown in \Fig{EcutEtacut}. 
For example, the energy threshold $\ecut$ decreases from the dwarf mass range ($\Mstar\la 10^{9}\Msun$ and $\Mv \la 10^{10.5}\Msun$) to sub-Galactic scale ($\Mstar\sim10^{9.5}\Msun$ and $\Mv\sim10^{11.5}\Msun$), and increases sharply at $\Mv\ga10^{12}\Msun$.

The trend is clearer with halo mass than with stellar mass.
At given halo mass and redshift, there is still significant halo-to-halo variance.
Hence, the kinematic thresholds are not constants, but exhibit significant systematic trends, as well as halo-to-halo variance at given mass and redshift.
Below, we speculate on possible causes of the median systematic trends in the halo mass.

The physical meaning of $\ecut$ is straightforward: it is a measurement of how bound is the dynamically cold part of the galaxy. 
The lower $\ecut$ is, the more settled the galaxy is, and the more extended the outskirt stellar halo is. 

Here we do not attempt at a full explanation of this V-shaped $\ecut$-mass relation at $\Mv<10^{12.5}\Msun$, but we believe that the low-mass trend is related to environmental effects. While only central galaxies are included in this analysis, lower mass systems are generally more affected by environmental processes and are more contaminated by backsplash satellites \citep{Wang23}. 
Environmental processes can drive $\ecut$ high, with tidal processes truncating the energy distribution so that the zero-potential surface shrinks \citep[e.g.,][]{Amorisco21}. 
We speculate that the $\ecut$ increase from $\Mv\sim10^{11.5}$ to $\Mv\sim10^{12.5}\Msun$
is related to changes in the accretion status of the haloes, but verification is beyond the scope of this study.
At the most massive end though ($\Mv\ga 10^{12.5}\Msun$), $\ecut$ seems to decrease with mass again, but we caution against overinterpreting this trend, as the massive end suffers from small-number statistics with the TNG50 sample. 

More interesting is the relation of the circularity threshold $\etacut$ with mass, as shown in the lower panels of \Fig{EcutEtacut}. 
Note that $\etacut$ is a gauge of how coherent is the AM of the thin disc, or simply put, how thin is the thin disc.
This should be distinguished from $\fthin$, the thin-disc mass fraction, because it can happen that a disc is thin but not significant in mass.
We can see that $\etacut$ increases with mass from $\sim 0.5$ at the dwarf scale to $\sim0.8$ at roughly the Milky-Way mass, $\Mstar\sim10^{10.5}\Msun$ or $\Mv\sim10^{12}\Msun$ \citep[e.g.,][]{Licquia15,Wimberly22}, and decreases towards higher masses.  
That is, the Milky-Way mass coincides with the mass scale at which coherent thin discs settle.
We also note that, while \cite{Du19} applied constant thresholds ($\ecut=-0.75$ for halo-bulge separation and $\etacut$>0.85 for cold disc), their thresholds are quite close to the median values we obtain here in the Milky-Way mass range.

\begin{figure*}	
\includegraphics[width=\textwidth]{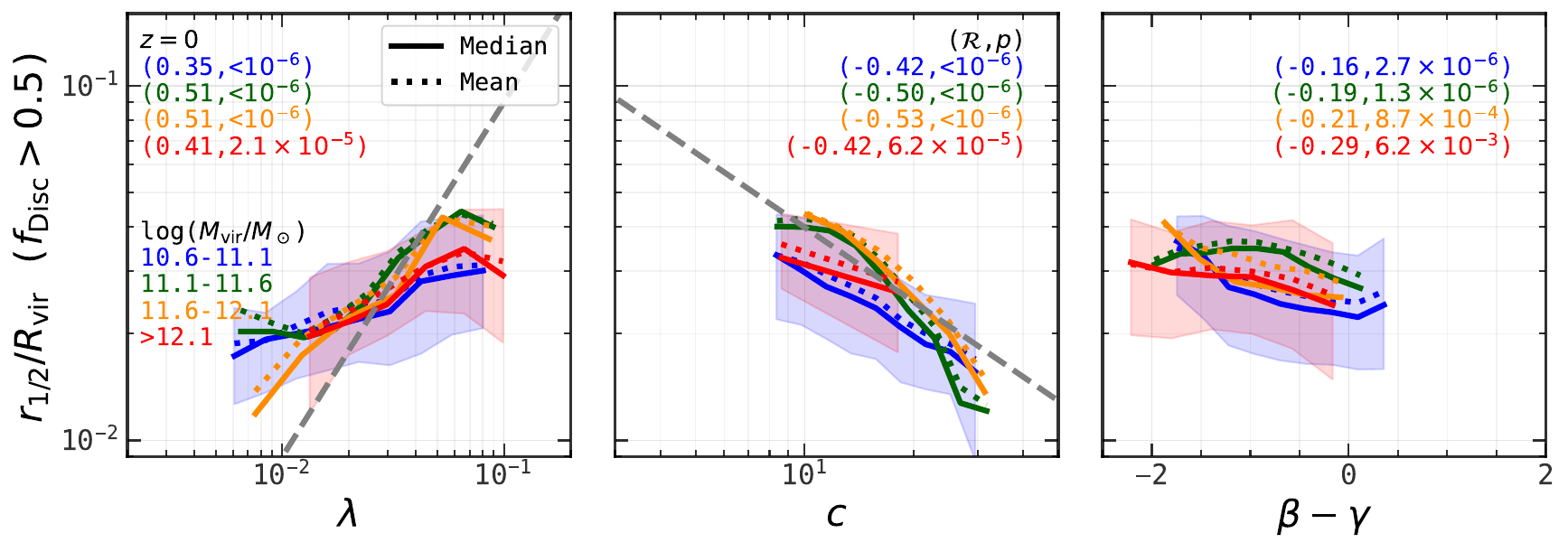}
    \caption{Galaxy compactness (galaxy half-stellar mass radius in units of host-halo virial radius) as a function of halo spin $\lambda$ (left), halo concentration $c$ (middle), and halo mass-assembly-history parameter $\beta-\gamma$ (right) for disc-dominated ($f_{\rm disc}>0.5$) central galaxies in different halo mass bins in the TNG50 simulations at z=0. 
Throughout the paper, we include all the galaxies that satisfy the binning strategy explicitly indicated in the legend, and do not apply additional trimming unless otherwise specified.
    The lines stand for the median or the average relations, as indicated, and the coloured bands indicate the 16th to 84th percentiles of the galaxy-compactness distribution at given fixed halo parameter (shown only for the lowest and highest mass bins to avoid making the plots too crowded). 
    Spearman correlation coefficients $\mathcal{R}$ and the corresponding $p$ values are quoted in the brackets in the form of $(\mathcal{R},p)$. 
    Gray dashed lines provide references of $\rhalf/\Rv\propto \lambda$ and slope $\rhalf/\Rv\propto c^{-0.7}$ in left and middle panels, respectively.}
    \label{fig:size}
\end{figure*}

\begin{figure*}	
\includegraphics[width=\textwidth]{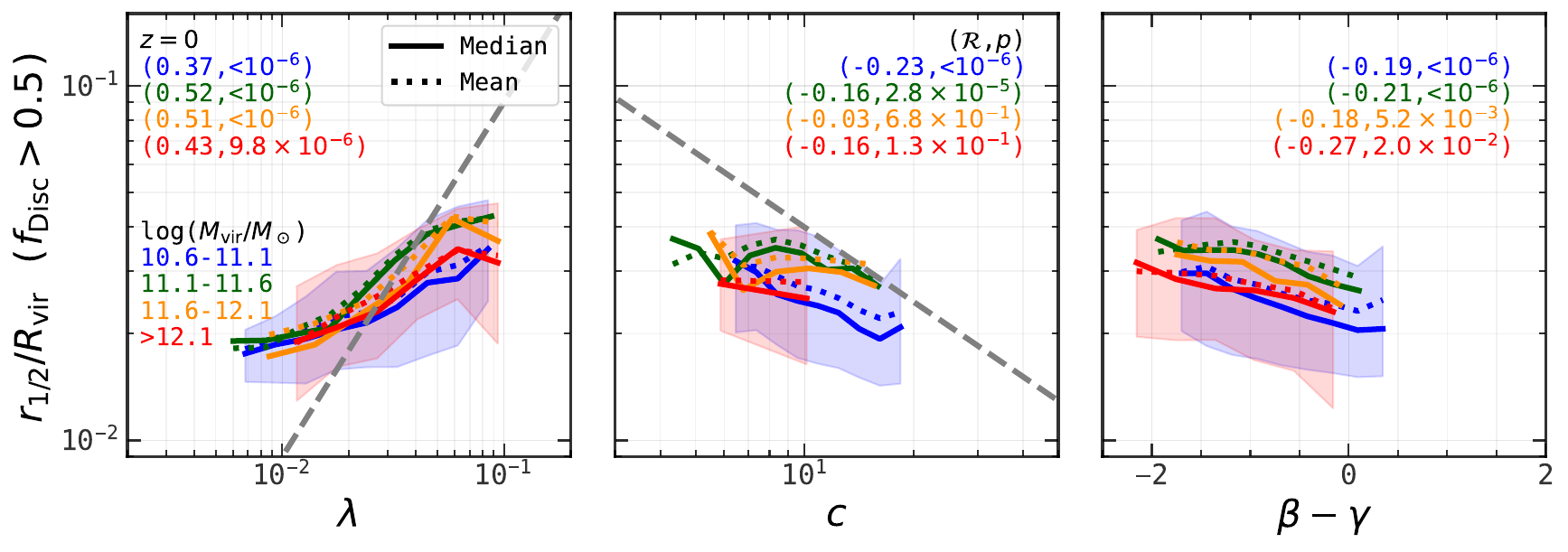}
    \caption{The same as \Fig{size} but with halo properties ($\Rv$, $\lambda$, $c$, and $\beta-\gamma$) measured in the DM-only simulations, TNG50-dark. 
    Results are shown for the galaxies with matched DM-only counterparts. 
}
    \label{fig:sizeDMO}
\end{figure*}

\begin{figure*}	
\includegraphics[width=\textwidth]{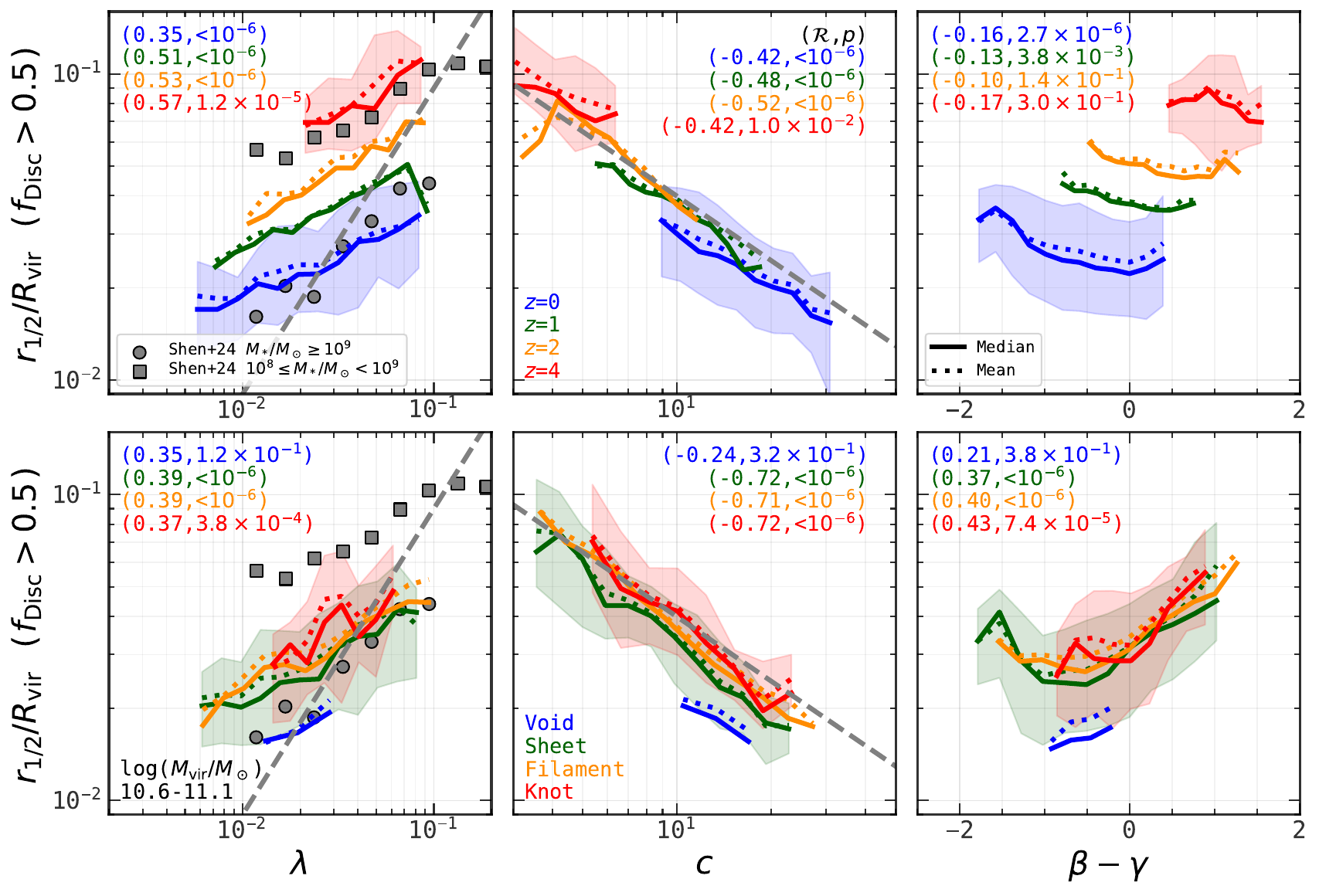}
    \caption{Similar to \Fig{size} but binning galaxies with given halo mass of $\Mv=10^{10.6-11.1}\Msun$ by redshifts (upper) and environments (lower panels). 
At each redshift in the upper panels, all the environments are included; similarly, for each environment in the lower panel, all the redshifts are included.
    With both halo mass and redshift fixed, the concentration dependence of the form of $\rhalf/\Rv\propto c^{-0.7}$ is clearer, and the correlation with halo accretion rate parameter ($\beta-\gamma$) is significant at lower redshifts.
    Overplotted for comparison in gray squares and circles are the high-redshift ($z\ga 5.5$) results of the THESAN-HR simulation, digitized and adapted from \citet[][see text]{Shen24} -- the spin dependence is similar.}
    \label{fig:sizefixedmass}
\end{figure*}

This characteristic scale of thin disc formation is more clear in terms of halo mass.
\cite{Dekel20} discovered with zoom-in hydro-cosmological simulations that galactic discs become stable only when the host galaxies reach $\Mv \ga 10^{11.5}\Msun$, and that this characteristic mass is nearly redshift-independent. 
Above this mass scale, discs start to settle, in the sense that they no longer flip in AM direction on short timescales and become more rotation supported.
Our findings here with a much larger statistical sample support this picture, in the sense that the halo mass at which the thin-disc threshold peaks is basically the same as that reported by \citeauthor{Dekel20} for disc settlement, and is similarly almost insensitive to redshift. 
The peak value $\etacut$ is redshift dependent though, but this simply reflects the fact that discs are generally less settled at higher $z$ when accretion is more intense and that the AM directions are constantly affected by the AM supply in the cosmic web. 
\citeauthor{Dekel20} attribute disc stabilization at $\Mv \ga 10^{11.5}\Msun$ to a process  they dub {\it gas-rich compaction}. 
This process gives rise to a compact star-forming bulge that stabilizes the disc \citep[see][for a thorough discussion]{Dekel20b,Lapiner23}. 
The exact mass scale at which this transition occurs may vary from one simulation to another, depending on various factors including feedback strength, and even in the same simulation suites, when investigating the characteristic mass from different perspectives, the detailed value may be slightly different. For example, \citeauthor{Lapiner23} quote a compaction halo mass of $\sim10^{12}\Msun$. Hence the mass scale should not be interpreted as a precision number.
While the TNG50 simulation lacks the resolution to fully resolve the compaction process, what we find here implies that the same behavior holds qualitatively.
The decrease of $\etacut$ at the galaxy-group scale likely reflects disc-thickening due to frequent heating by mergers and secular evolution.
To consolidate these speculations is beyond the scope of current work.
Here, we conclude that simplistic decomposition with constant $\etacut$ of $\sim0.7$ would result in higher mass fraction of thin discs for Milky-Way-mass galaxies, and lower in thin discs for dwarfs. 

We emphasize that thin and thick in this study are in a relative sense. That is, even if a disc is thickened by dynamical heating, or at higher redshift when discs are generally kinematically hotter, we can still identify a thinner component as long as two intersecting Gaussian components exist. The idea of treating thin and thick in a relative sense is in keeping with the findings from previous observational and theoretical studies. For example, \citet{Wisnioski15} and \citet{Kretschmer22} demonstrate that the ratio between rotational velocity and velocity dispersion, $v_{\rm rot}/\sigma$, can vary systematically across redshifts, decreasing from $\gtrsim10$ for nearby spirals to $\sim 3$ for high-$z$ discs. 
If one assumes a constant $v_{\rm rot}/\sigma$ threshold for selecting discs, then a threshold suitable for high-$z$ discs would yield an unrealistic  high fraction of galaxies classified as discs in the local Universe. 
Although here we focus on circularity as the kinematic metric instead of $v_{\rm rot}/\sigma$, the rationale is similar, that the metric needs to be adaptive to reflect general situations at an epoch.

\begin{figure*}	
\includegraphics[width=\textwidth]{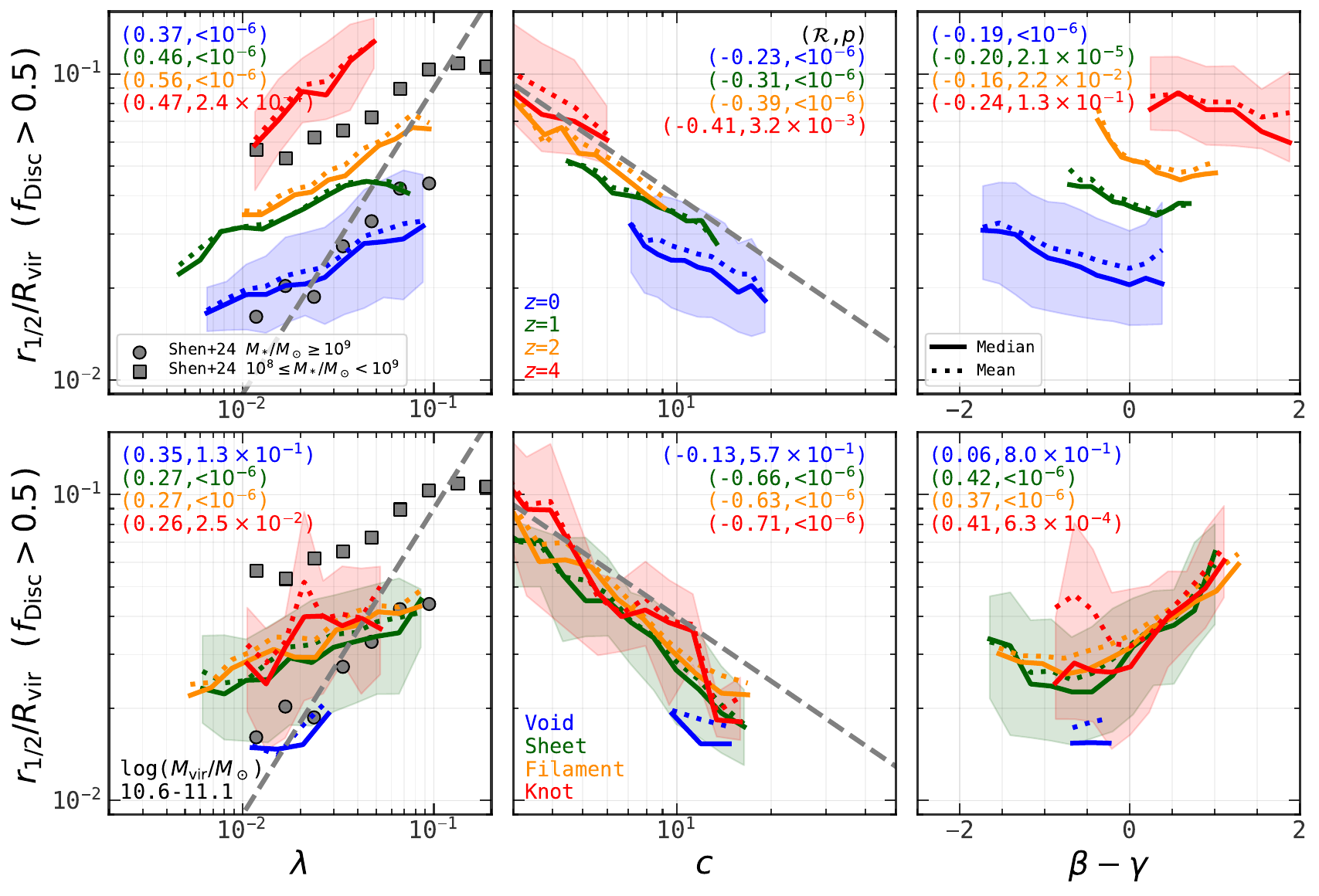}
    \caption{The same as \Fig{sizefixedmass} but with halo properties measured in the DM-only simulations, TNG50-dark. 
    Results are shown for the galaxies with matched DM-only counterparts. 
}
    \label{fig:sizeDMOfixedmass}
\end{figure*}

%--------------------------------------------------
%--------------------------------------------------
\section{Galaxy morphology and halo structure}\label{sec:result}

With different morphological components identified, we study how galaxy morphology is related to the structures of their hosting DM haloes.
First, we identify the leading factors affecting the sizes of disc-dominated galaxies (\se{size}). 
Second, we explore the dependence of the mass ratios of different morphological components on halo properties (\se{ratios}).
%We leave a more thorough investigation of galaxy-morphology-halo-structure connections to Paper II, where we apply machine-learning algorithms to further characterize the correlations. 

%--------------------------------------------------
\subsection{Disc-size predictor revisited}\label{sec:size}

The size of disc galaxies is a key topic in galaxy formation theory, yet there is no consensus on which DM halo properties beyond virial mass (or radius) best correlate with disc size, or whether DM properties alone can accurately predict it---a central question in semi-analytical galaxy evolution models. Assuming AM conservation, \citet{Mo98} shows that disc size scales with the halo spin parameter via
\be\label{rhalfMMW98}
\rhalf \simeq f_j \lambda \Rv,
\ee
where $f_j$ is the AM retention factor. This links disc size to halo spin, providing a convenient framework for galaxy size modeling. However, \citet{Jiang19} reveals, through zoom-in cosmological simulations, that halo spin shows weak correlation with galaxy AM and limited predictive power for size, especially at 
$z\ga 1$. 
Instead, they propose an empirical relation involving halo concentration:
\be\label{rhalfJiang19}
\rhalf \simeq 0.02 (1+z)^{-0.2} c_{10}^{-0.7} \Rv,
\ee
where $c_{10}$ is the concentration normalized to a value of 10 for Milky Way-like haloes at $z\sim0$. This relation partially captures redshift dependence, contrasting with the \citeauthor{Mo98} model, which has no explicit redshift dependence. 
Both models agree on a leading-order scaling, 
$\rhalf\sim0.02\Rv$, consistent with abundance-matching results \citep{Kravtsov13, Somerville18}. 
Since halo radius (mass) is the dominant determinant of galaxy size, we focus on the ratio 
$\rhalf/\Rv$, termed the {\it galaxy compactness}, and explore its dependence on secondary halo parameters.

In what follows, we present the correlations between galaxy compactness and halo structural parameters that we find to be relevant, namely, the halo spin ($\lambda$), the halo concentration ($c$), and the proxy for accretion rate ($\beta-\gamma$). We focus on disc-dominated galaxies, defined as those with a total disc fraction (thin+thick) larger than 50 per cent. We adopt different binning strategies by mass and by redshift (or environment), respectively, in \Figs{size}-\ref{fig:sizeDMO} and \Figs{sizefixedmass}-\ref{fig:sizeDMOfixedmass}, in order to reveal potential systematic trends in the size-halo correlations.

We first study the correlations at $z=0$ for different halo mass bins. 
As shown in \Fig{size}, the spin parameter $\lambda$ is indeed positively correlated with $\rhalf/\Rv$, almost insensitive to the mass range of interest. However, the average relation is shallower than $\rhalf/\Rv\propto \lambda$. 
With a Spearman correlation coefficient of $\mathcal{R}\sim 0.3-0.5$, the correlation of galaxy size with spin is notably stronger than that found in \citet{Jiang19}, which is based on zoom-in simulations of higher resolution than TNG50. 
We note that even in the two zoom-in suites that \citeauthor{Jiang19} analyzed, the lower-resolution NIHAO sample exhibits somewhat stronger correlation with halo spin than that in the higher-resolution VELA sample, especially at low $z$ (see Fig.A1 therein).
These imply that numerical resolution might play a role in connecting the AM of the galaxy and that of the host halo. 
Different stellar feedback prescriptions amongst these simulations could also impact the AM connection, as well as other aspects of galaxy-halo structural connections. 
The results shown here only reflect the TNG50 simulation and are not necessarily generic. For example, depending on the detailed implementation of feedback in a simulation, the fraction of gas turned into stars may sample the total halo gas very differently, and the correlation between galaxy size with halo spin may be different in simulations where the coupling of the gas and the feedback energy is modeled differently. 
For example, \cite{Yang23} compared the relations between galaxy compactness and halo spin from the TNG family of simulations and that from the EAGLE simulations, and found the correlations in TNG family to be stronger, and speculatively attributed the difference to the stronger baryon cycles in the TNG runs.

There is almost no redshift dependence on the correlation strength with halo spin, as can be seen from the upper left panel of \Fig{sizefixedmass}. \citet{Shen24} studied galaxy size in the THESAN-HR simulations in the reionization epoch, and actually found a very similar level of correlation strength with halo spin, as overlayed in \Fig{sizefixedmass} for comparison. 
\footnote{Their original result was shown between $\rhalf$ and $\lambda$, and in order to adapt their result for comparison in our plots of $\rhalf/\Rv$, we have estimated the median $\Rv$ for their two stellar mass bins by using the galaxies of corresponding masses in TNG50, assuming that they have similar halo masses as in the THESAN-HR simulation. 
This assumption is justified as we focus on the correlation strength with spin rather than the detailed normalization of galaxy compactness.}
The THESAN-HR simulation features on-the-fly radiative transfer (RT), and adopts otherwise similar subgrid prescriptions to TNG50. 
Hence, we infer that real-time RT does not play a crucial role in establishing or eliminating the spin dependence, but other aspects of feedback might, given that both TNG50 and THESAN-HR exhibit similarly strong $\lambda$ dependence, stronger than that in the NIHAO and VELA suites.  

With the larger sample of TNG50, we verify the anti-correlation between disc compactness and halo concentration first revealed with zoom-in simulations. 
With a Spearman coefficient of $\mathcal{R}\sim-0.45$, the trend is actually quite close to the $c^{-0.7}$ scaling as previously found by \cite{Jiang19} imposing a simple power law dependence.

\begin{figure*}	
\includegraphics[width=0.7\textwidth]{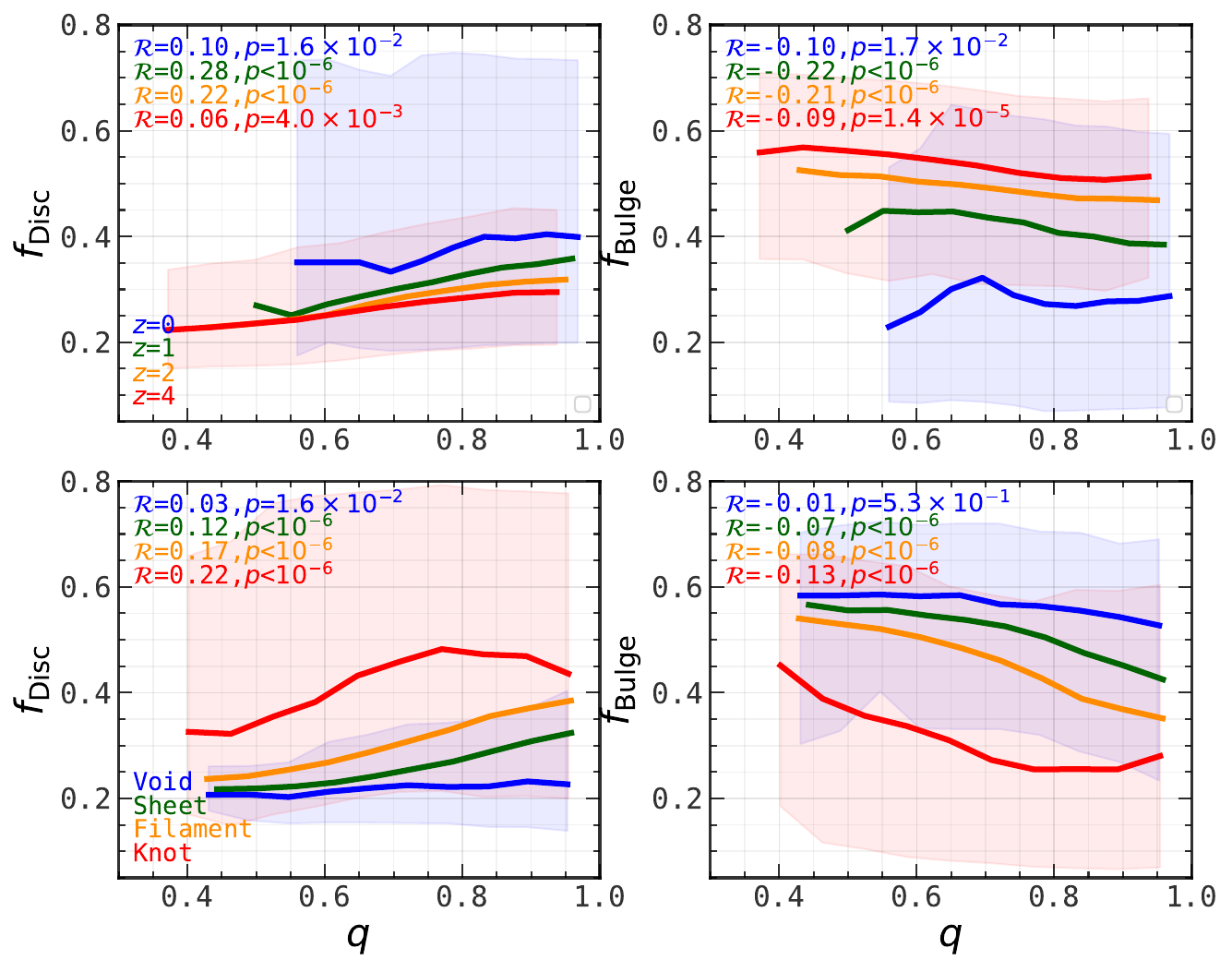}
    \caption{The mass fraction of discs (left) and bulges (right) as a function of halo shape $q$ for central galaxies at different redshifts (upper) or in different environments (lower). Here we focus on the median trend, but note that the scatter in $f_X$ at fixed halo axis ratio $q$ is actually large, comparable to the vertical plotting range. The Spearman coefficient $\mathcal{R}$ and corresponding $p$ values are quoted.}
    \label{fig:fraction_q}
\end{figure*}

A negative $c$ dependence may arise from adiabatic contraction of DM haloes in response to the inhabitant disc potentials. 
Semi-anlaytical or semi-empirical modelers who base their galaxy prescriptions on DMO simulations may question the predicting power of the halo structural parameters if these dependences arise from baryonic effects such as adiabatic contraction.
To test this scenario, we use the DM-only (DMO) run and focus on haloes with matched counterparts between the DMO run and the full-physics run.
We measure their stellar properties in the hydrodynamical run, and the DM properties including $\Rv$, $\lambda$, and $c$ are from the matched haloes in the DMO run.
The results are shown in \Fig{sizeDMO}, where we can see that the $c$ dependence is indeed weakened, with the Spearman correlation coefficient dropped significantly to $\sim -0.2$ from the $\mathcal{R}\sim-0.45$ as in the hydrodynamical run, implying that it indeed arises partially from halo response.
This is particularly obvious in the more massive bins.
However, there is still a statistically significant negative correlation between $\rhalf/\Rv$ and the DMO concentration for dwarf haloes and sub-Galactic haloes of the mass range $\Mv=10^{10.6-11.6}\Msun$.

We try to further clarify the trend with $c$ by keeping both the halo mass and redshift fixed, as shown in \Fig{sizeDMOfixedmass}.
We choose the mass bin of $\Mv=10^{10.6-11.1}\Msun$ in order to have both decent statistics and a mass range that is sufficiently narrow to rule out contamination from mass dependence.  
Clearly, the same concentration scaling basically holds for haloes in the fixed dwarf-halo mass bin and at each fixed redshift.
We do not extend this investigation of fixed mass and fixed redshift to higher mass bins because the TNG50 box would be too small to give sufficient statistics for massive halos at high $z$.
We conclude that, while the concentration dependence is weaker than the spin dependence, it is not purely a consequence of halo response so that $c$ from DMO simulations still have predicting power on galaxy size , at least for dwarf systems. 

Besides halo spin and concentration, the MAH of the halo also affects disc size, as shown in the last column of \Figs{size}-\ref{fig:sizeDMOfixedmass}.
Here, we have parameterized the MAH by the parameter combination $\beta-\gamma$, which, as introduced in \se{HaloQuantities}, reflects the overall shape of the MAH, and approximates the accretion rate at redshift zero. 
In the more actively accreting haloes (which have more negative $\beta-\gamma$), galaxies are more extended. 
This correlation is rather weak, with a Spearman coefficient of $\cal{R}\sim$ -0.15 and $\simeq -0.2$ in the hydrodynamical run and DMO run respectively, but seems to be significant at least at the lower redshifts as indicated by the small $p$ values there. 

Finally, we explore the impact of environment on these correlations by dividing the galaxies according to their cosmic-web identities, as shown in the lower panels of \Fig{sizefixedmass}-\ref{fig:sizeDMOfixedmass}. 
There is no clear-cut trends with cosmic-web environments.
Galaxies in denser environments seem to be slightly more extended than those in voids, but we caution against interpreting this due to the low statistics in voids. 

%\begin{figure}
%  \centering
%  \subfloat{
%    \includegraphics[width=\columnwidth]{Figures/roRvirSph_q.pdf}
%  }
%  \hfill % add desired spacing
%  \subfloat{
%    \includegraphics[width=\columnwidth]{Figures/roRvirSph_q_DMO.pdf}
%  }
%  \caption{}
%  \label{}
%\end{figure}

%Above, we have focused on disc-dominated galaxies, for bulge dominated systems ($f_\mathrm{sph}$) we find that the dependences of galaxy compactness $\rhalf/\Rv$ on concentration or MAH are qualitatively the same as those for the discs.
%There is no correlation with the halo spin parameter, as expected, since spheroidal systems arises from mergers, gas-rich compaction, or disc instabilties \citep[e.g.,][]{Lapiner23}, all of which erase the information of the AM of the DM halo from the baryonic end products.  
%There is a weak but noticeable correlation with the 3D shape of DM haloes though, that the spheroidal compactness $\rhalf/\Rv$ shows a negative correlation with the axis ratio $q$, for dwarf haloes and Milky-Way-mass systems ($\Mv\la 10^{12}\Msun$) 

%--------------------------------------------------
\subsection{Mass fractions of morphological components}\label{sec:ratios}

We find the mass ratios of different morphological components to also correlate with host-halo properties.
In this study, we focus on the disk fractions and the circularity threshold.
Notably, the disc fraction $\fdisc$ correlates positively with the 3D axis ratio $q$ of DM haloes, as shown in the left column of \Fig{fraction_q}. 
The axis ratio $q$ is an indicator of how relaxed a system is \citep{Schneider12,Bonamigo15,Menker22}, and we have verified  that other triaxiality indicators (i.e., shape parameters involving combinations of $q$ and $p$) or relaxedness indicators (e.g., $T/|U|$, the ratio between kinetic and potential energy) yields qualitatively similar trends. 
Hence, relaxed and rounder systems ($q\sim 1$) tend to host more significant discs. 
The bulge fraction $\fbulge$, on the other hand, correlates negatively with halo shape $q$, in the sense that the bulge fraction is higher in more elongated systems. 
Both trends are weak and of large scatter, but are robust against binning the samples by redshift or cosmic-web environment. The redshift and environmental trends are both straightforward to understand, that lower-redshift and knot galaxies are more disc dominated, coherent with the picture that disc formation needs more settled environments.

\begin{figure}	
\includegraphics[width=0.5\textwidth]{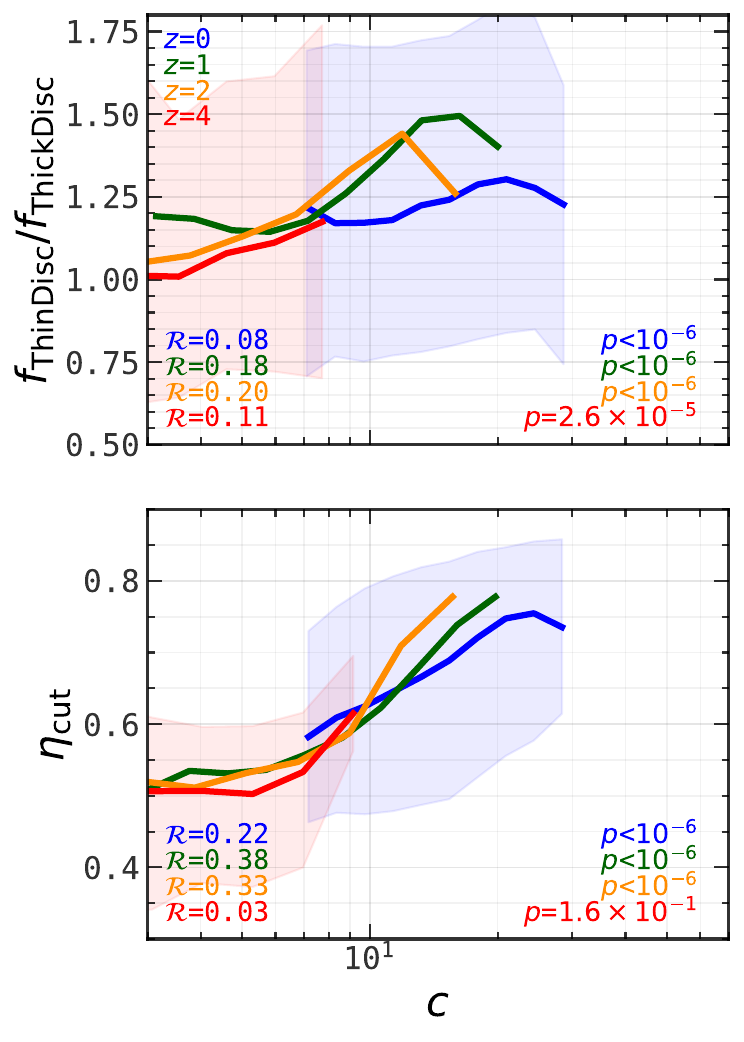}
    \caption{The median mass ratio between thin and thick discs $\fthin/\fthick$ (upper), and the median circularity threshold $\etacut$ (lower) as a function of the halo concentration $c$ at different redshifts.
    }
    \label{fig:MAH_c_vs_etacut_fD}
\end{figure}

\bigskip
The mass ratio between the thin and thick discs, $\fthin/\fthick$, as well as the circularity threshold $\etacut$ used for separating them, also depends on the DM-halo properties.
This is shown in \Fig{MAH_c_vs_etacut_fD}.
Here we reiterate that $\etacut$ is subtly different from the mass ratio $\fthin/\fthick$, as it measures how thin (i.e., how coherent in AM) the thin disc is, but not necessarily the mass dominance.
The lower the concentration, the less developed the thin disc in terms of both the mass ratio and the threshold $\etacut$. 
Since $c$ is related to the accretion status of a system \citep[e.g.,][]{Ludlow13} and low $c$ is related to recent major mergers \citep{Wang20}, this trend basically reiterates that the development of a significant thin disc prefers a stable, relaxed halo.

%--------------------------------------------------
%--------------------------------------------------
\section{Discussion}\label{sec:discussion}

In this section, we quantify the impact of different decomposition methods on the morphological mass fractions, and address the environmental dependences of $\fdisc$ and $\fbulge$ at different redshifts. 

%--------------------------------------------------
\subsection{Comparison with the Zana et al. (2022) method}\label{sec:Zana}

\begin{figure*}	
\includegraphics[width=0.7\textwidth]{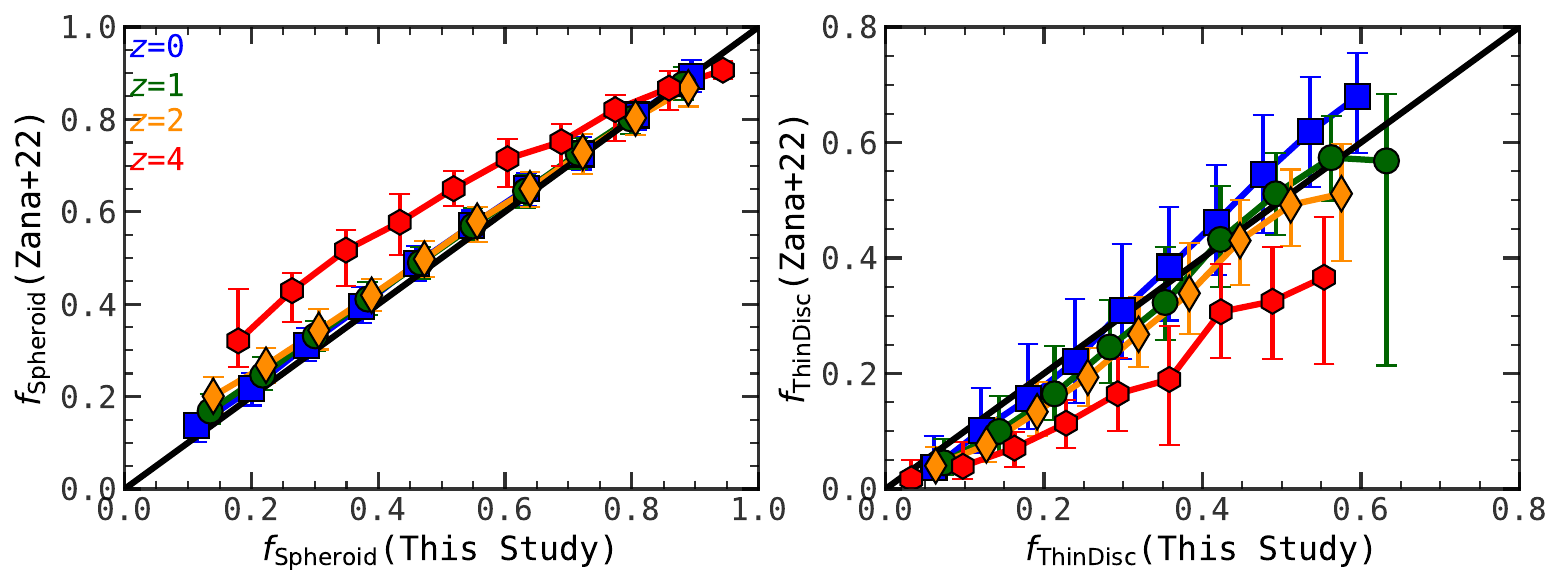}
    \caption{Comparison with the \protect\cite{Zana22} method, regarding the spheroidal mass fraction (left) and the thin-disc mass fraction (right) at different redshifts. 
    The symbols are the medians with the error bars indicating the 16th and 84th percentiles.}
    \label{fig:comparison_zana}
\end{figure*}

Our new method inherits the framework of \cite{Zana22}, with the separation of the high-energy and low-energy components basically the same, except for some fine-tuning details accounting for the higher mass resolution of the TNG50 sample.
Our key improvement here is the running circularity threshold for thin disc as revealed by the GMM algorithm, as opposed to their constant $\etacut$ of 0.7. 
We have demonstrated in \Fig{MW-method} and \Figs{Massive-method}-\ref{fig:sph-method} that this method gives robust identification of thin and thick discs in the circularity-polarity-energy space even when the one-point circularity distribution exhibits no bimodality. 
\Fig{comparison_zana} shows the mass fraction of spheroids (halo+bulge) and thin discs at different redshifts, comparing the results from this work and from the method of \citeauthor{Zana22}. 
\footnote{
For the result of \citeauthor{Zana22}, we use their results publicly available at \href{https://www.tng-project.org/data/docs/specifications/\#sec5t}{https://www.tng-project.org/data/docs/specifications/\#sec5t}.
}
First, since our energy criteria are similar, the mass fraction of spheroids are similar. 
There is a slight difference at $z\leq 2$ and a larger difference at $z=4$ -- this is mainly because \citeauthor{Zana22} did not exclude wind particles while we do, and the fraction of wind particles is higher at higher redshifts. 
The wind particles are short lived and are controlled by hydrodynamics and feedback prescriptions instead of gravitational dynamics, so it is reasonable to exclude them when focusing on the morphology of long-lived stars.
This leads to a lower mass fraction of spheroids, more so at higher redshifts when star formation and stellar feedback were more intense. 

Second, our new method yields thin discs that are more massive at higher redshift. 
This behavior can be anticipated from what is shown in \Fig{EcutEtacut}, that most low-$z$ galaxies have $\etacut > 0.7$, except for the low-mass dwarfs of $\Ms <10^{8.8}\Msun$, while most high-$z$ galaxies have $\etacut < 0.7$. 
Similar comparisons could be made for different mass bins or cosmic-web identities, and all lead to the conclusion that the running circularity threshold yields non-monotonic differences with respect to the constant one of 0.7.

\subsection{Comparison with Yu et al. (2021) model}\label{sec:Yu}

\begin{figure*}	
\includegraphics[width=\textwidth]{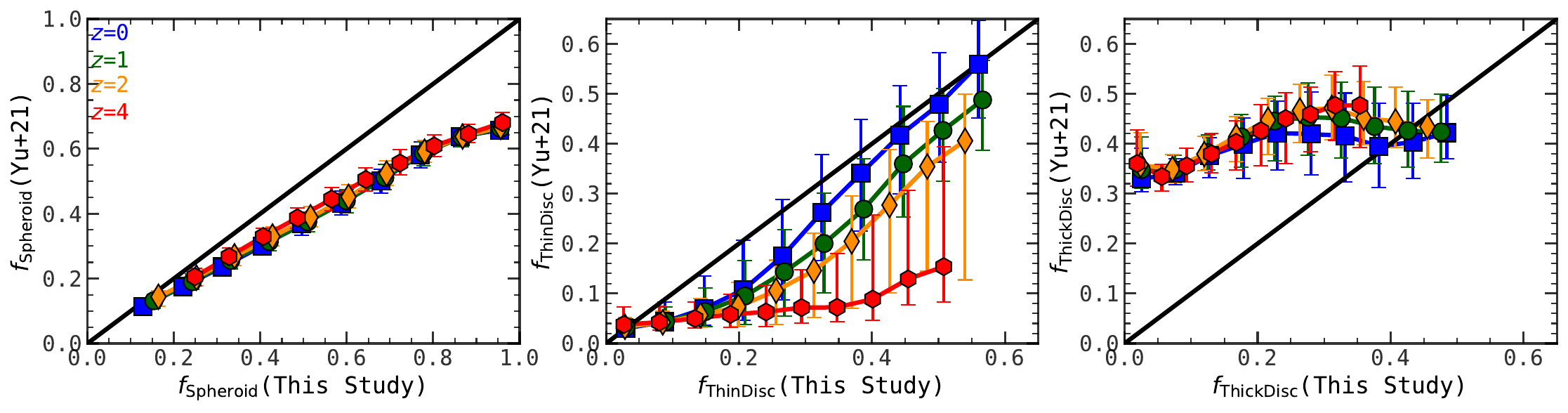}
    \caption{Comparison with the \protect\cite{Yu21,Yu23} method based on constant circularity thresholds, regarding the mass fractions of spheroidal components (left), thin discs (middle), and thick discs (right). The symbols represent the medians, with the error bars indicating the 16th and 84th percetiles. The simple method significantly leads to lower mass fraction in thin discs at high-$z$.}
    \label{fig:comparison_Yu}
\end{figure*}

\Fig{comparison_Yu} compares the our method to that of \citet{Yu21,Yu23} which uses constant circularity cuts for both bulge and discs. 
Recall that the \citeauthor{Yu23} method identifies the stellar particles with $\eta< 0.2$ as spheroids, and assigns those with $\eta>0.8$ and $0.2<\eta<0.8$ to the thin and thick discs, respectively. 
This is a representative case of the most commonly used simple kinematic decomposition scheme in the literature \citep[e.g.,][]{Tacchella19, Sotillo-Ramos23}. 
Their method yields a consistently smaller spheroidal fraction, and gives a lower mass fraction of thin discs at all redshifts.\footnote{We follow their method regarding the circularity thresholds, but we note that the radius range of computing the galaxy AM is slightly different: \citeauthor{Yu23} use 0.1$\Rv$ while we use $5\rhalf$. Here we focus on the effect of the running threshold so assume so we use $5\rhalf$ for both results.}
The offset is particularly strong at high redshift.
Given the difference that could be introduced by different decomposition schemes, we caution against blindly comparing morphological mass fractions obtained from different studies, even when the studies may seem to base on similar principles.
This again can be understood via \Fig{EcutEtacut}, which shows that $\etacut$ is almost always smaller than 0.8, especially for high-redshift galaxies. 
Since the simple decomposition scheme yields lower mass fraction in both thin discs and spheroids, the thick disc mass fraction is generally higher.

We not only caution against using simple decomposition schemes based on constant kinematics thresholds for morphological comparisons with JWST observations, we also caution against using blindly using our methods for such purposes. 
The observed morphologies at high redshift are usually based on photometry rather than kinematics, and to characterize the systematics between kinematics morphology in simulations and optical morphology in observations is beyond the scope of current study. 
In a follow-up work, we generate mock JWST observations and quantify such systematics.

\subsection{Environment dependence}\label{sec:environment}

\begin{figure*}	
\includegraphics[width=\textwidth]{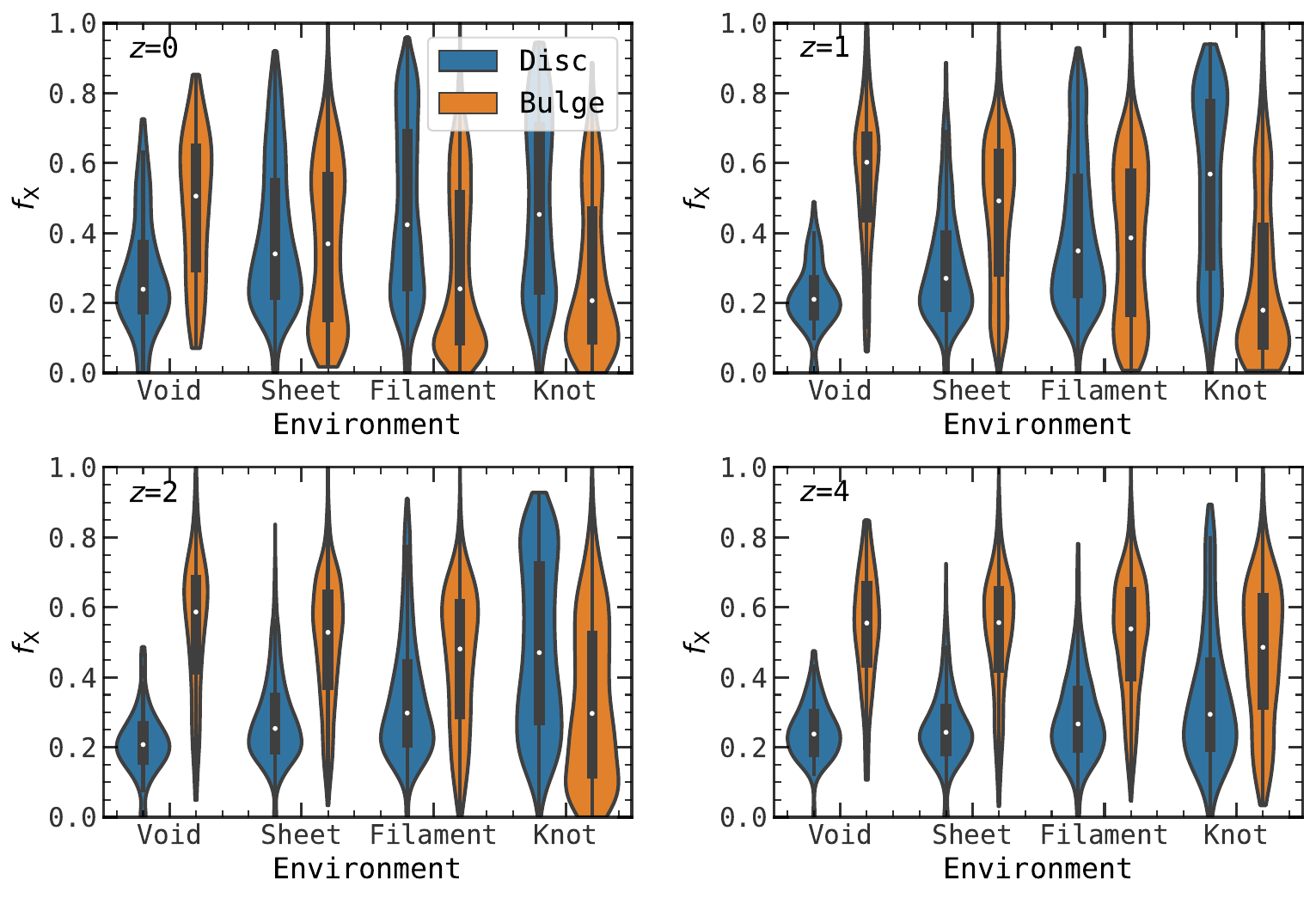}
    \caption{`Violin' plots showing the distributions of the mass-fraction distributions of disc (blue) and bulge (orange) as a function of environment classification, at different redshifts. 
    The white points stand for the median values while the shapes of `violin' stand for the normalized distributions of mass fraction. 
    The central thick bars represent the 25th and 75th percentiles and the thin lines indicate the maximum and minimun values.}
    \label{fig:environment}
\end{figure*}

Above, we presented briefly the cosmic-web dependence of galaxy morphology. Notably, the cosmic-web environment has a weak impact on disc size at fixed halo mass, and the mass fractions of bulges and discs as functions of halo shape also varies in different cosmic-web classifications.
Here we elaborate on these trends, comparing in \Fig{environment} the distributions of disc mass fraction $\fdisc$ and bulge mass fraction $\fbulge$ in different environments at fixed redshifts.
First, there is a general behavior across redshifts, that the average bulge fraction decreases towards denser environments while the disc fraction increases from voids to knots.
This is particularly clear at intermediate redshifts of $z\sim 1-2$: in knots, the median disc fraction $\langle \fdisc \rangle$ is higher than the median bulge fraction $\langle \fbulge \rangle$, but in voids, $\langle \fdisc \rangle$ drops to $\sim 0.2$ while $\langle \fbulge \rangle$ increases to $\sim 0.6$.
At $z=4$, the bulge fraction is always higher than the disc fraction on average, and it is reasonable to speculate that this is the case at even earlier epochs during reionization.  
This trend is in keeping with the scenario that disc development requires a stable environment, in the sense that central haloes in low-$z$ knots are the most relaxed systems (except for the perturbed backsplash haloes).

Second, the $\fdisc$ and $\fbulge$ distributions are basically unimodal in the least dense environments of voids and sheets, but exhibit bimodality in knots (at $z\la 2$) and filaments (at $z=0$). 
The distributions are more extended with larger range of 25th-75th percentiles in denser environments. 
Hence, galaxy morphological diversity is affected by environment. 
The unimodal distributions at the highest redshift suggests that it takes time for environmental effects to accumulate, or that the intense cosmic accretion at early times washed out these effects to some extent. 
The unimodal $\fdisc$ distributions in voids is relatively narrow and peak at $\fdisc\sim0.2$, indicating that massive discs can hardly develop in isolated environments or at the borders of cosmic knots\footnote{In fact, we note that the void class and knot class wane and wax when one dials the somewhat arbitrary thresholds of the eigenvalues $\lambdath$ of the deformation tensor. Hence, some of the void galaxies are just outside the border of knots. }, this likely manifests limited gas supply in such environments because disc growth requires continuous gas supply in the first place.
We have verified that these trends hold if we further keep the halo mass fixed, but opted to show the full sample for better statistics. 

%--------------------------------------------------
%--------------------------------------------------

\section{Conclusions}\label{sec:conclusion}

In this work, we introduce a new morphological decomposition scheme for galaxies in cosmological simulations using stellar kinematics. 
Similar to \citet{Zana22}, our new decomposition scheme first detects the most prominent local minimum of the energy distribution of star particles and thus breaks a galaxy into higher-energy and lower-energy components -- the random-motion supported subsets of which are identified as the stellar halo and the bulge, respectively. 
It then classifies the remaining rotationally supported stars into two groups in the 3D space spanned by specific energy ($\eb$), circularity ($\eta\equiv\jz/\jc$), and polarity ($\epsilon \equiv \jp/\jc$), using the Gaussian-Mixture-Models algorithm, and identifies the circularity threshold $\etacut$ for thin and thick discs as the intersection point of the $\eta$ distributions of the two groups. 

We have applied this method to the TNG50 simulations and briefly revisited the connection between galaxy morphology and DM halo structure, considering halo spin, concentration, mass-assembly history (MAH), as well as the location in the cosmic web. 
We leave more systematic analyses to a future work of this series, but highlight the following take-away messages. Regarding morphological decomposition and morphological fractions, we find that --  

\begin{itemize}[leftmargin=*]

\item The GMM algorithm reveals a circularity threshold for thin disc ($\etacut$) that shows galaxy-to-galaxy variance, and systematic variation with halo mass and redshift.
The median $\etacut$ increases with mass from $\simeq 0.5$ at the dwarf regime ($\Mv\sim10^{10}\Msun$) to the peak value of $\simeq 0.8$ at the Milky-Way ($\Mv\sim10^{12}\Msun$) scale, with the exact values slightly higher at lower $z$, and decreases towards higher masses. 
The energy threshold $\ecut$ also exhibits significant halo-to-halo variance and systematic mass and redshift dependence, with the minimum occurring at slightly below the Milky-Way mass ($\Mv\sim10^{11.5}\Msun$).
These mass trends can be seen with both stellar mass and halo mass, but are sharpened with halo mass. 
The characteristic halo mass at which $\etacut$ peaks or $\ecut$ minimizes barely depends on redshift. 
These all hint at the DM halo playing crucial role in regulating the morphology of inhabitant galaxies, and are in synergy with the theoretical picture generalized from zoom-in cosmological simulations \citep{Dekel20} that gas-rich compaction preludes disc development, and that there is a characteristic halo mass $\Mv\sim10^{11.5}\Msun$ at which the regime transition occurs.  

\item The constant circularity threshold of $\sim$ 0.7 or 0.8 widely used in the literature \citep[e.g.,][]{Yu21,Yu23,Tacchella19, Sotillo-Ramos23} is arguably oversimplified and would result in dramatically different disc fractions -- smaller thin discs and heavier thick discs -- from what we find.
This is particularly alarming at high redshift $z\ga4$ and thus highlights the danger of blindly comparing simulation results or comparing simulations to observations if the decomposition is based on kinematics thresholds. 
\end{itemize}

For the structural connections between galaxies and their host haloes, we find that -- 

\begin{itemize}[leftmargin=*]

\item The half-mass size of disc-dominated galaxies, or rather, the compactness of the stellar-disc mass distribution with respect to the virial radius of the host halo, $\rhalf/\Rv$, is positively correlated with the DM halo spin parameter $\lambda$ and negatively correlated with halo concentration $c$. 
The correlation with spin is noticeably stronger than that reported earlier for zoom-in simulations \citep{Jiang19}, but is weaker than than the linear scaling as in the seminal \citet{Mo98} model of disc size. 
The correlation strength is comparable to the high-$z$ results of the THESAN-HR simulation \citep{Shen24}, which inherits the TNG subgrid physics but adds on-the-fly radiative transfer. 

\item The concentration dependence of galaxy compactness is similar to that in \cite{Jiang19}, captured roughly by a $c^{-0.7}$ scaling. 
The concentration dependence is not merely a consequence of the halo contraction in response to the baryonic potential, because it holds even if the concentration is measured from the DM-only simulation for the galaxies with matched counterparts between the full-physics run and the DM-only run. The $c$ dependence is indeed weakened with the DM-only measurements, but it at least exists for dwarf haloes.

\item The sizes of disc-dominated galaxies depend on redshift such that $\rhalf/\Rv$ on average increases from $\sim 0.02$ at $z=0$ to $\sim 0.07$ at $z= 4$ or the end of reionization.
Newly revealed in this study is that disc compactness also correlates with the mass accretion rate of the DM halo.
When halo mass and redshift are fixed, galaxies are more extended in more actively accreting haloes. 
Specifically, the correlation between $\rhalf/\Rv$ and the halo MAH parameter $\beta-\gamma$ is weak but robust at redshift up to 2, with a Spearman correlation coefficient of $\mathcal{R}\simeq -0.2$.

\item Besides disc size, the mass ratio between thin and thick discs is lower for lower-concentration halos, in keeping with the interpretation that thin disc development needs stable halo conditions given that $c$ is an indicator of dynamical status.

\item The disc mass fraction $\fdisc$ exhibits positive correlation with the 3D axis ratio $q$ of host halo. 
With $q$ as an indicator of how relaxed the system is, this reiterates the point that disc development requires stable relaxed haloes.
The bulge mass fraction $\fbulge$ shows the opposite trend with the halo shape $q$,
suggesting that bulge build-up is related to mergers or other processes that disturb the host halo. 
Both trends are stronger in denser cosmic-web environments. 

\end{itemize}

Overall, with our new GMM-aided morphological decomposition and the exhaustive halo-structure measurements, we conclude that the morphological diversity of galaxies indeed contains useful information about their host halo. 
These correlations are however complex and in many cases rather weak, posing a challenge to the whole business of quantifying galaxy-halo structural connections. 
We caution that this study is more about laying out a framework for such kind of analysis. The detailed results reported here are specific to the TNG50 simulation and thus could be strongly dependent on the TNG sub-grid physics. 
We emphasize though that our method is public and is implemented in a modular way so it can be easily adapted to other simulations.
We leave a more detailed study and tentative physical explanations to Paper II, and encourage the community to adapt our methods to other cosmological simulations to further tackle this interesting but challenging issue.

%--------------------------------------------------
%--------------------------------------------------
\section*{Acknowledgements}

FJ and JL thank Joel Primack, Sandra Faber, and Cedric Lacey for helpful discussions, and appreciate the Tsinghua Astrophysics High-Performance Computing platform for providing computational resources for this work. 
AD is partly supported by the Israel Science Foundation grant 861/20. 
JL acknowledges the support of the UK Science and Technology Facilities Council (STFC) studentship (ST/Y509346/1).
LCH was supported by the National Science Foundation of China (11991052, 12011540375, 12233001), the National Key R\&D Program of China (2022YFF0503401), and the China Manned Space Project (CMS-CSST-2021-A04, CMS-CSST-2021-A06).
We thank the anonymous referee for comprehensive comments.

%%%%%%%%%%%%%%%%%%%%%%%%%%%%%%%%%%%%%%%%%%%%%%%%%%
\section*{Data Availability}

We make our pipeline for morphological decomposition and halo-property measurements publicly available at \href{https://github.com/JinningLianggithub/MorphDecom}{https://github.com/JinningLianggithub/MorphDecom}.
The catalogs of TNG50 galaxies with detailed morphological measurements and fittings are available upon reasonable requests and will be available when Paper II is published.  

%%%%%%%%%%%%%%%%%%%% REFERENCES %%%%%%%%%%%%%%%%%%

% The best way to enter references is to use BibTeX:

\bibliographystyle{mnras}
\bibliography{PaperI}

% Alternatively you could enter them by hand, like this:
% This method is tedious and prone to error if you have lots of references
%\begin{thebibliography}{99}
%\bibitem[\protect\citeauthoryear{Author}{2012}]{Author2012}
%Author A.~N., 2013, Journal of Improbable Astronomy, 1, 1
%\bibitem[\protect\citeauthoryear{Others}{2013}]{Others2013}
%Others S., 2012, Journal of Interesting Stuff, 17, 198
%\end{thebibliography}

%%%%%%%%%%%%%%%%%%%%%%%%%%%%%%%%%%%%%%%%%%%%%%%%%%

%%%%%%%%%%%%%%%%% APPENDICES %%%%%%%%%%%%%%%%%%%%%

\appendix

%--------------------------------------------------
%--------------------------------------------------

\section{Visualization for More Examples}\label{app:vis}
\begin{figure*}	
\includegraphics[width=0.8\textwidth]{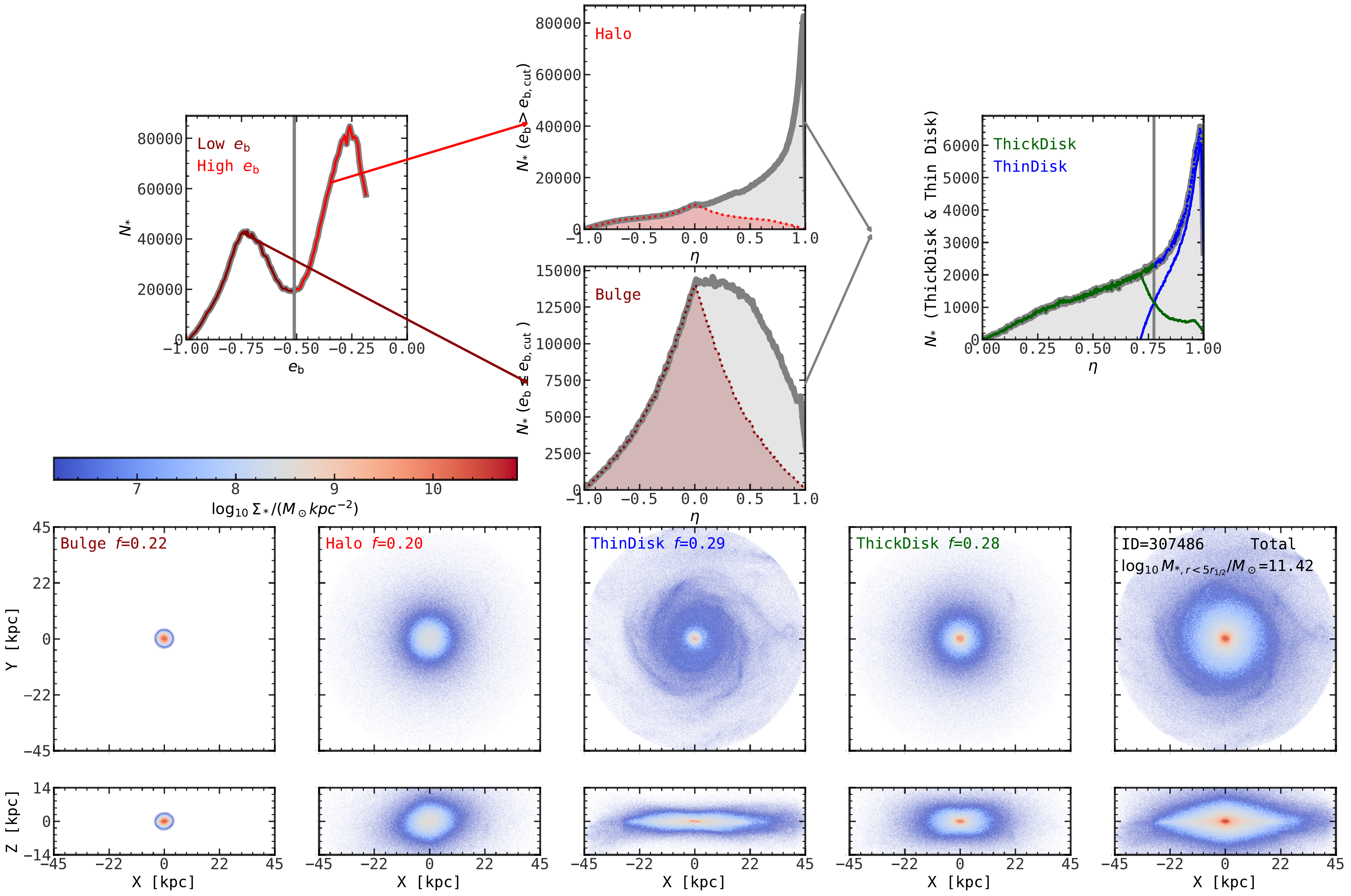}
    \caption{Similar to Figure \ref{fig:MW-method} but for massive galaxy at $z=0$.}
    \label{fig:Massive-method}
\end{figure*}

\begin{figure*}	
\includegraphics[width=0.8\textwidth]{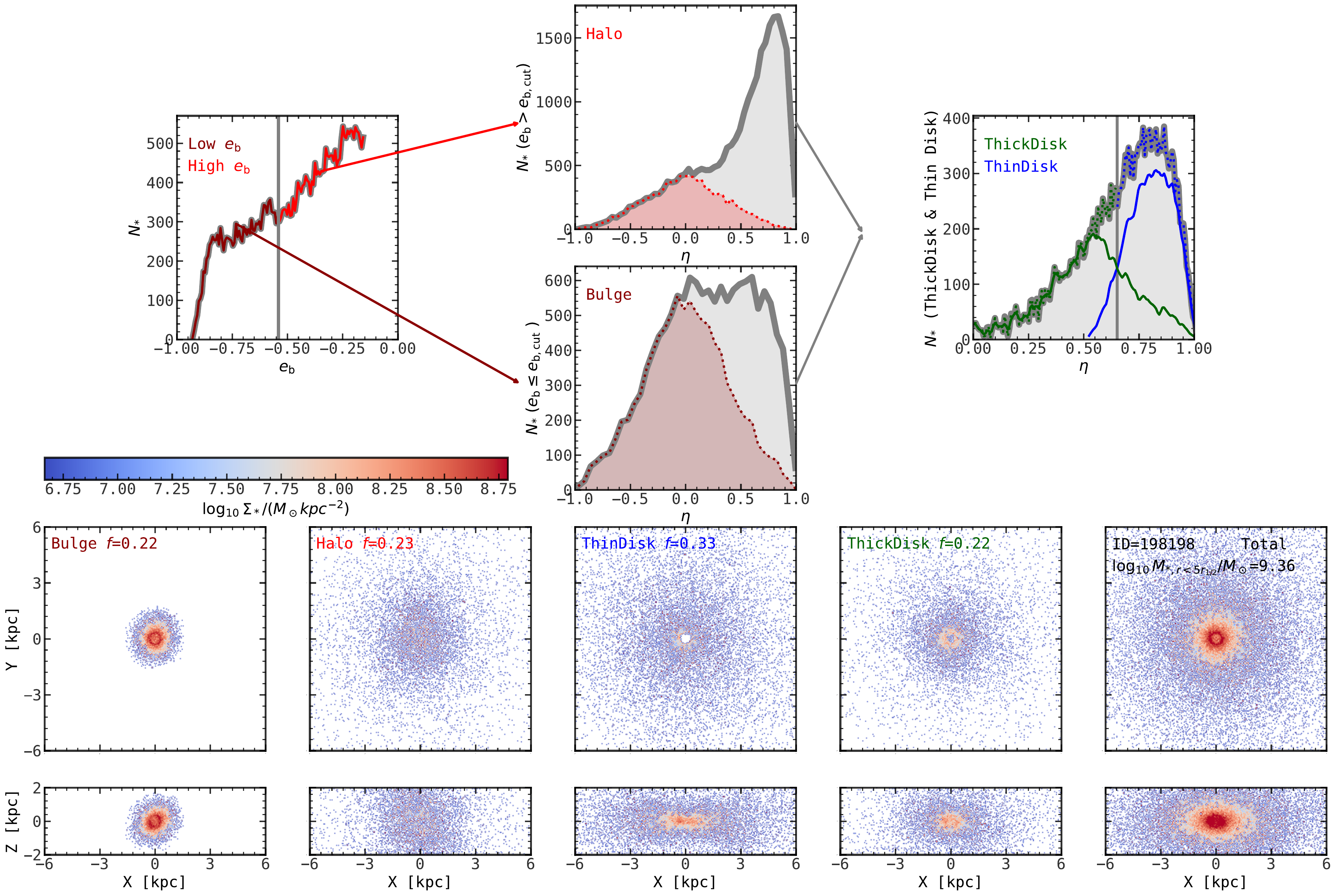}
    \caption{Similar to Figure \ref{fig:MW-method} but for dwarf galaxy at $z=0$.} 
    \label{fig:Dwarf-method}
\end{figure*}

\begin{figure*}	
\includegraphics[width=0.8\textwidth]{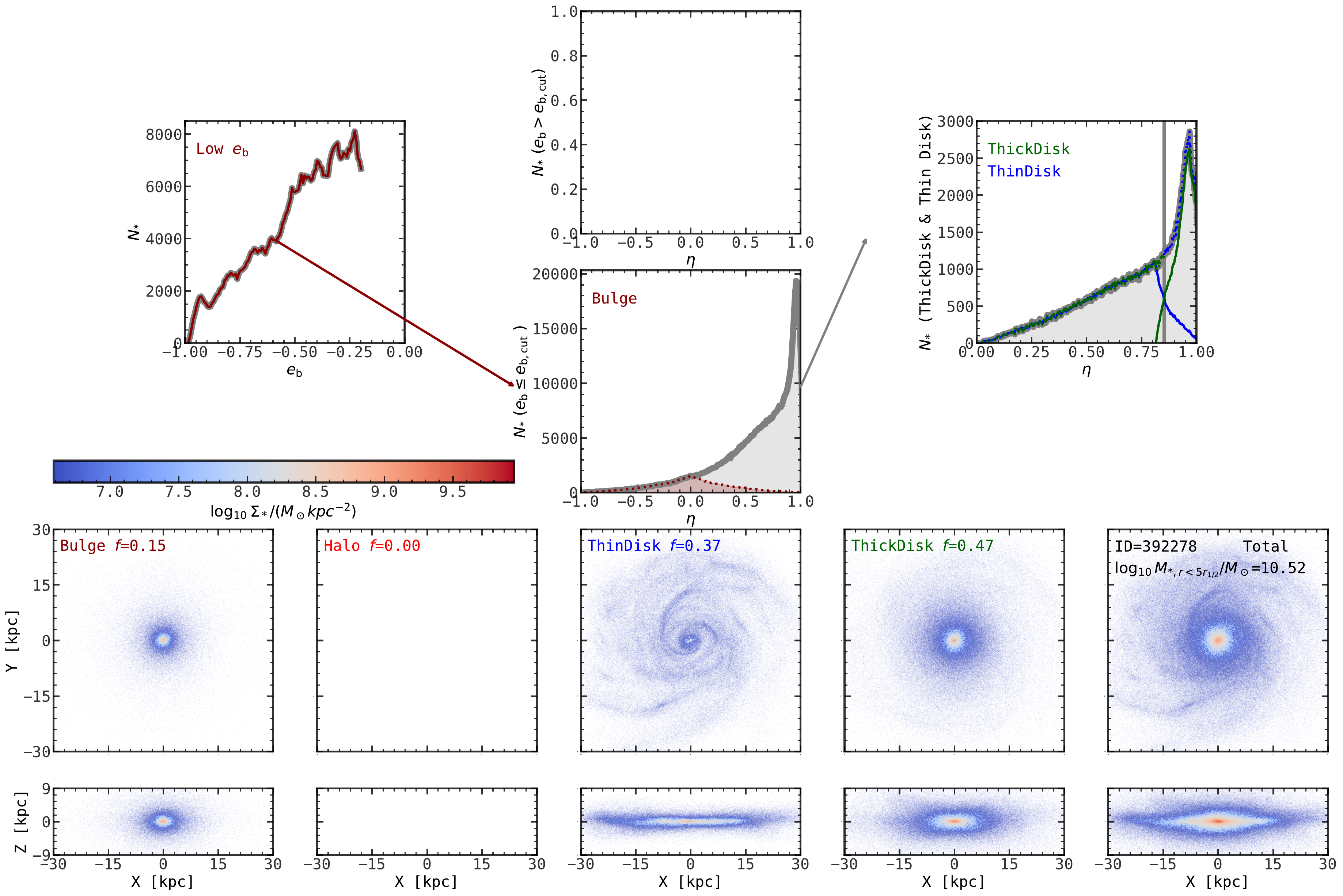}
    \caption{Similar to Figure \ref{fig:MW-method} but for disc-dominated galaxy at $z=0$. Note that in such galaxies, it is possible that there is no energy minima to be found, which leads to this galaxy having no stellar halo}
    \label{fig:disk-method}
\end{figure*}

\begin{figure*}	
\includegraphics[width=0.8\textwidth]{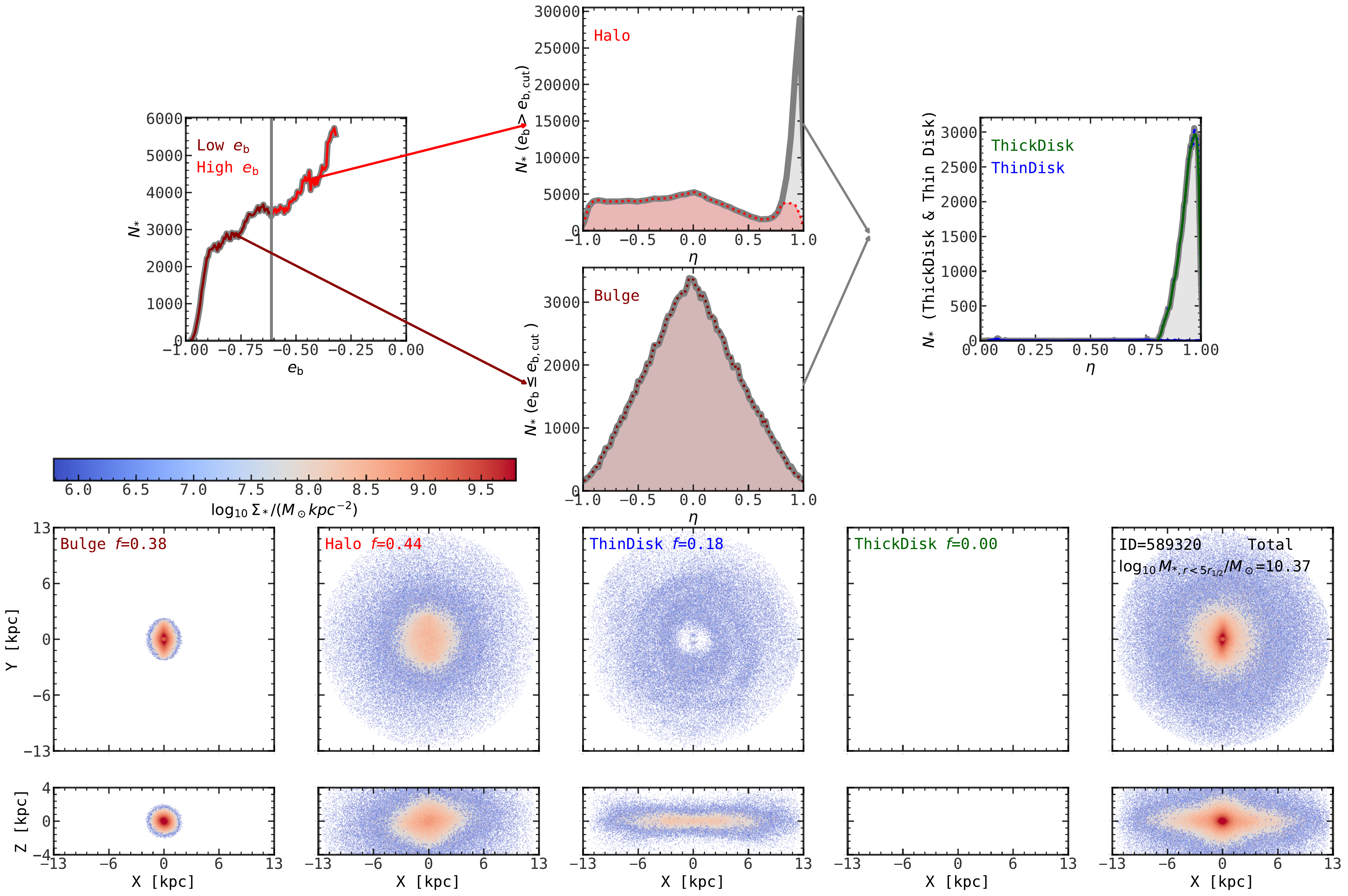}
    \caption{Similar to Figure \ref{fig:MW-method} but for spheroidal-dominated galaxy at $z=0$. Note that in such galaxies, it is possible that there is no circularity threshold to be found, which leads to this galaxy having only one disc (thin disc).}
    \label{fig:sph-method}
\end{figure*}

In this Appendix, we show more examples of morphological decomposition for different types of galaxies using method. 
\Fig{Massive-method} shows a massive galaxy at redshift 0 with comparable mass fractions for the four components. 
\Fig{Dwarf-method} presents a dwarf galaxy. 
\Fig{disk-method} and \Fig{sph-method} show extreme cases of a disc with marginal stellar halo component and a bulge-dominated system, respectively. 
These are stress tests though, as we have verified that, with our method, bulgeless pure discs (defined as $\fdisc > 0.9$) are rare, of percent level in TNG50 at $z=0$ and negligible at $z\ga 2$, and the fraction of systems with no stellar halo ($\fhalo < 0.01$) is uniformly rare ($\la 6\%$) across redshifts. 

%--------------------------------------------------
%--------------------------------------------------

\section{Algorithm for detecting the energy threshold}\label{app:EnergyThreshold}

First, we divide the $\eb$ distribution into 25 bins ranging from its minimum to the 90th percentile. 
We then go through each bin $i$ and register all the local minima in the distribution that satisfy the following conditions: (A) $N_{i-1}$ > $N_{i}$ and $N_{i+1}$ $\geq$ $N_{i}$ or $N_{i-1}$ $\geq$ $N_{i}$ and $N_{i+1}$ > $N_{i}$ with $N_{i}$ the number of particles in the $i$-th bin, (B) $N_{i-2}$ > $N_{i}$ and $N_{i+2}$ > $N_{i}$, and (C) $\sum_{j}^{j>i}N_{j} > N_{\rm min}$ where
$N_{\rm min} = \text{max}(N_{\rm crit},0.01N_{\star})$, with $N_\star$ the total number of bound stellar particles considered and $N_{\rm crit}$ a critical number depending on $N_{\star}$: $N_{\rm crit}=1000$ when $N_{\star}\geq 10^4$, $N_{\rm crit}=100$ when $10^3\leq N_{\star}< 10^4$, $N_{\rm crit}=10$ when $N_{\star}\leq 10^3$. 
\citeauthor{Zana22} set $N_{\rm crit}$=1000 for all galaxies irrespective of the mass scale. 
Our adaptive scheme yields more robust local minima detection for dwarf galaxies, which are common in the TNG50 box. 
If no minimum is detected with the three conditions above, the search is extended to the full $\eb$ distribution. 
If there is still no detection, we postulate that there is no high-energy component and thus no stellar halo in this galaxy (and the only random-motion supported component is registered as the stellar bulge). 
If, instead, multiple minima are detected, we iteratively repeat the search with half the bin size, and require that the distance from the newly detected minima to a previously found one to be smaller than $3\Delta e_{\rm bin}$, with $\Delta e_{\rm bin}$ the bin size at the current iteration step. 
We stop searching either when a single minimum is left, or when the number of bins reaches a maximum value, which, following \citeauthor{Zana22}, is set to be the integer part of $\sqrt{N_\star/2}$ in the range between 80 and 400. 
To avoid the minima which are very close to galactic centres, we require the mass ratio between the lower-energy component and the higher-energy component to be larger than 0.05, and also require $\ecut$ candidates to be larger than -0.9. 
After trimming, if there are still multiple candidates, we adopt the lowest one as $\ecut$. 
To further refine the value of $\ecut$ so that it is the $\eb$ of a specific stellar particle, we iteratively take the median $\eb$ of the stellar particles within the interval centered on the minimum found in the last iteration until the interval width reaches the minimal value corresponding to the maximum number of bins. 

%--------------------------------------------------
%--------------------------------------------------

\section{Gaussian Mixture Model}\label{app:GMM}
Mixture models can be used to describe a multi-dimensional distribution $p(\bm x)$ by combination of $K$ base distributions. In GMM, base distribution is chosen as Gaussian distribution. When the sample data is multivariate, Gaussian distribution function takes form of 
\be
\mathcal{N}(\bm x|\bm \mu,\bm \Sigma)=\frac{1}{(2\pi)^{\frac{D}{2}}|\bm \Sigma|^\frac{1}{2}}\text{exp}\left(-\frac{(\bm x-\bm \mu)^T\bm \Sigma^{-1}(\bm x-\bm \mu)}{2}\right)
\ee
where $\bm x$ is data points, $\bm \mu$ is the expectation of data, $\bm \Sigma$ is the covariance of data and $D$ is the dimension of data.

When we combine a finite number of $K$ Gaussian distribution function to express the distribution function of data, one can get
\be
p(\bm x|\bm \theta)=\sum_{k=1}^K\pi_k \mathcal{N}(\bm x|\bm \mu,\bm \Sigma)
\ee
where $\pi_k$ is the $k$th mixture weights which can be treat as the probability of $N(\bm x|\bm \mu,\bm \Sigma)$ and satisfy
\be
0\leq \pi_k \leq 1, \sum_{k=1}^K\pi_k=1
\ee
where we define $\bm \theta=\{\bm \mu_k, \bm \Sigma_k, \bm \pi_k : k=1,\cdots, K\}$. To get $\bm \theta$ with the maximal probability, one needs to do optimization via maximum likelihood.

Unlike a single Gaussian distribution, the likelihood of a mixture model cannot be expressed as closed-form. Therefore, one can only get $\bm \theta$ by iteration.

Assume we are given a dataset $\bm\chi=\{\bm x_1,\cdots,\bm x_N\}$, where $\bm x_{n}$, $n=1,\cdots, N$ is a sample in this dataset. Log-likelihood can be expressed as 
\be
\log \left(\bm \chi | \bm \theta\right)=\sum_{n=1}^N\log\sum_{k=1}^K\pi_k \mathcal{N}(\bm x_n|\bm \mu_k,\bm \Sigma_k)
\ee
One can calculate the parameter $\bm{\theta}$ by the following steps:

1. Initialize $\bm \mu_{k}$, $\bm \Sigma_k$ and $\pi_k$

2. Calculate $\mathcal{N}(\bm x_n|\bm \mu_k,\bm \Sigma_k)$ and $p(\bm x|\bm \theta)$

3. Update $\bm \mu_{k}$, $\bm \Sigma_k$ and $\pi_k$

4. Repeat 2 and 3, until the total likelihood $\log \left(\bm \chi | \bm \theta\right)$ converges or when it reach threshold. 

Then one gets $K$ determined Gaussian components to express distribution of data. 

%1. Initialize $\bm \mu_{k}$, $\bm \Sigma_k$ and $\pi_k$

%2. Evaluate responsibilities $r_{nk}$ for each data sample $\bm x_n$ by
%\be
%r_{nk}=\frac{\pi_k \mathcal{N}(\bm x_n|\bm \mu_k,\bm \Sigma_k)}{\sum_{j=1}^K\pi_j \mathcal{N}(\bm x_n|\bm \mu_j,\bm \Sigma_j)}
%\ee

%3. Estimate updated $\bm \mu_{k}$, $\bm \Sigma_k$ and $\pi_k$ by
%\be
%\bm \mu_{k}=\frac{1}{N_k}\sum_{n=1}^N r_{nk}\bm x_n
%\ee
%\be
%\bm \Sigma_k=\frac{1}{N_k}\sum_{n=1}^N r_{nk}\left(\bm x_n-\bm \mu_k\right)\left(\bm x_n-\bm \mu_k\right)^T
%\ee
%\be
%\pi_k=\frac{N_k}{N}
%\ee
%where $N_k=\sum_{n=1}^N r_{nk}$

%Using updated parameters one can re-calculate log-likelihood and stop when it reach threshold. Then one gets $K$ determined Gaussian components to express distribution of data.Mixture models can be used to describe a distribution $p(x)$ by combination of $K$ base distributions. 

\section{GMM components in three dimensional space}\label{app:3DGMM}
\begin{figure}	
\includegraphics[width=\columnwidth]{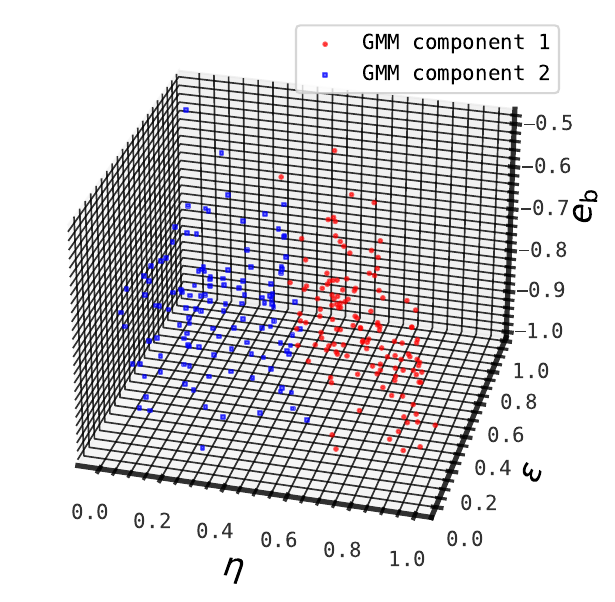}
    \caption{GMM visualization in three dimensional space for disky stars in a dwarf galaxy at redshift 0 (ID 775066). The red circles are GMM component 1, most of which will be identified as a thin disc while blue squares are GMM component 2, most of which will be identified as a thick disc.}
    \label{fig:3DGMM}
\end{figure}

In Section \ref{sec:decomposition}, we use GMM method to separate disky stars into thin and thick disc. The marginal distribution for $\eta$ of two gaussian components can be seen in \Fig{MW-method} and \Fig{Massive-method} - \Fig{sph-method}. While one can see that the marginal distribution is overlapped, we note that the distribution in three dimensional space, i.e. $\eta$-$\epsilon$-$e_{\rm b}$ space cannot be overlapped. To prove this, we use an dwarf central galaxy (ID 775066) at redshift 0 as an example since smaller number of particles is better for visualization. As one can see in \Fig{3DGMM}, two gaussian components represented by blue and red points are separated clearly. This indicate that GMM can successfully distinguish two groups.

%%%%%%%%%%%%%%%%%%%%%%%%%%%%%%%%%%%%%%%%%%%%%%%%%%

% Don't change these lines
\bsp	% typesetting comment
\label{lastpage}
\end{document}